\documentclass[11pt]{article}
\pdfoutput=1

\usepackage{aas_macros,amsmath,cite,graphicx,mathtools,setspace,subfigure}
\usepackage[margin=1in,letterpaper]{geometry}
\usepackage[colorlinks=true]{hyperref}
\usepackage[affil-it]{authblk}

\numberwithin{equation}{section}
\newtagform{brackets}{[}{]}

\newcommand{\ab}[1]{\left|#1\right|}
\newcommand{\br}[1]{\left[#1\right]}
\newcommand{\cu}[1]{\left\{#1\right\}}
\newcommand{\pa}[1]{\left( #1 \right)}

\newcommand{\ed}{\,\mathrm{d}}
\newcommand{\pd}{\,\partial}
\renewcommand{\L}{\mathcal{L}}

\begin{document}

\title{\Huge Force-Free Foliations}
\date{}
\author[1]{Geoffrey Comp\`{e}re\thanks{gcompere@ulb.ac.be}}
\author[2]{Samuel E. Gralla\thanks{sgralla@email.arizona.edu}}
\author[3]{Alexandru Lupsasca\thanks{lupsasca@fas.harvard.edu}}
\affil[1]{\small
Universit\'{e} Libre de Bruxelles and International Solvay Institutes\\
Brussels, CP 231 B-1050, Belgium}
\affil[2]{\small
Department of Physics, University of Arizona\\
Tucson, AZ 85721, USA}
\affil[3]{\small
Center for the Fundamental Laws of Nature, Harvard University\\
Cambridge, MA 02138, USA}

\maketitle

\begin{abstract}
Electromagnetic field configurations with vanishing Lorentz force density are known as \textit{force-free} and appear in terrestrial, space, and astrophysical plasmas.  We explore a general method for finding such configurations based on formulating equations for the field \textit{lines} rather than the field itself.  The basic object becomes a foliation of spacetime or, in the stationary axisymmetric case, of the half-plane.  We use this approach to find some new stationary and axisymmetric solutions, one of which could represent a rotating plasma vortex near a magnetic null point.
\end{abstract}

\vfill\pagebreak

\tableofcontents

\section{Introduction}

A Maxwell field $F_{\mu\nu}$ satisfying $F_{\mu\nu}J^\nu=0$, where $J^\mu=\nabla_\nu F^{\mu\nu}$ is the four-current, is known as \textit{force-free}.  Force-free fields are ubiquitous in nature: they can be found in the laboratory \cite{Gray2013}, in the solar corona \cite{Wiegelmann2012}, near neutron stars \cite{Goldreich1969,Michel1973}, and near black holes \cite{Blandford1977}.  After decades of study by plasma physicists, solar physicists, and astrophysicists, there has recently been new interest in the force-free equations from the general relativity and high energy physics communities \cite{Palenzuela2010,Neilsen2011,Palenzuela2011,Moesta2012,Lehner2012,Alic2012,Paschalidis2013a,Paschalidis2013b,Brennan2013,Ruiz2014,Brennan2014,Gralla2014,Lupsasca2014,Zhang2014,Lupsasca2015,Yang2014,Wang2014,Ponce2014,Gralla2015a,Gralla2015b,Jacobson2015,Gralla2016a,Gralla2016b,Yang2015,Zhang2015,Compere2016,Wang2015,Yang2016,Gralla2016c,Gralla2016d,Carrasco2016}.  As a simple nonlinear system with a nevertheless intricate structure, these equations are of mathematical interest in their own right.

The force-free equations are written compactly in terms of the Maxwell two-form $F_{\mu\nu}$ as
\begin{align}
\label{eq:FFE}
	F_{\mu\nu}\nabla_\rho F^{\rho\nu}=0,\qquad\nabla_{[\mu}F_{\nu\rho]}=0.
\end{align}
The first equation is the force-free condition, while the second is the statement that the form is closed (no magnetic monopoles).  (Here, $\nabla_\mu$ is compatible with the spacetime metric $g_{\mu\nu}$.)  Vacuum solutions with $J^\mu=\nabla_\nu F^{\mu\nu}=0$ comprise a trivial subset on which the equations become linear.  Provided that $J^\mu\neq0$, Eqs.~\eqref{eq:FFE} imply that the two-form $F_{\mu\nu}$ is simple or, equivalently, \textit{degenerate},
\begin{align}
	F_{[\mu\nu}F_{\rho\sigma]}=0.
\end{align}
Reviews of the rich physics of force-free fields may be found in Refs.~\cite{Wiegelmann2012,Gralla2014}.  In this paper, we concentrate on the mathematical problem of finding solutions to the nonlinear system \eqref{eq:FFE}.

The technique we pursue is motivated by a beautiful observation due to Carter \cite{Carter1979}: degenerate, closed two-forms define a \textit{foliation} of spacetime into two-surfaces (see also Refs.~\cite{Uchida1997,Gralla2014}).  These surfaces are spanned by the vectors $V^\mu$ such that $F_{\mu\nu}V^\nu=0$, and are interpreted as worldsheets of magnetic field lines in the magnetically dominated case $F_{\mu\nu}F^{\mu\nu}=2\pa{B^2-E^2}>0$ of physical interest.  Since force-free fields are degenerate, each force-free field determines a foliation.  The converse is not true in general, but if a foliation does determine a force-free field (in the magnetic case), then that field is unique (see Appendix \ref{app:Foliation}).  This means that the force-free condition can be reexpressed as a condition on foliations.  Thus, one passes from the field to the field \textit{lines} as the fundamental variable.

One can hope that such a reformulation will lead to new insights and results.  In this paper, we perform a version of this reformulation specialized to stationary, axisymmetric, force-free fields.  Such fields are characterized by three scalars defined on the ``poloidal (half-)plane'' spanned by the cylindrical radius $\rho>0$ and height $z$: the flux function $\psi(\rho,z)$, polar current $I(\psi)$, and field angular velocity $\Omega(\psi)$.  Many of the most interesting exact solutions have been found by guessing a common functional dependence on some scalar $u$, i.e., by making the ansatz $\psi=\psi(u)$, $I=I(u)$ and $\Omega=\Omega(u)$.  One then examines the force-free condition to see if an associated solution exists or not.  Previously, this has been done on a case-by-case basis, but it would be desirable to have a more systematic method for checking whether a function $u$ is admissible or not, i.e., to pass to $u$ as the basic variable.  Since the level sets of $u$ are the poloidal projections of the magnetic field lines, this is a version of the foliation strategy outlined above.

We are able to eliminate $\psi$ and $I(\psi)$ in favor of the foliation representative $u$, but in general, $\Omega(u)$ remains present.  We give the equation in coordinate form as well as in terms of geometric invariants of the foliation, and ultimately work in a general stationary, axisymmetric (circular) spacetime.  The equation is most useful in the case $\Omega=0$ (or more generally, constant $\Omega$), where it becomes a single ``foliation condition'' on $u$.  There is a large gauge redundancy in this description, since two functions $u$ with parallel gradient have the same level sets and hence correspond to the same foliation.  This makes the foliation equation appear more complicated than the original force-free equation (at least when written in coordinate form), but it also means that it has many more solutions, thereby making them easier to guess.  We can use the foliation equation as a consistency condition to check whether or not a force-free solution exists.  If the check is successful, then it is straightforward to reconstruct the solution.

Having this foliation condition enables automation of the guesswork by computer algebra programs.  In Sec.~\ref{sec:ForceFreeMagneticFields}, we describe a simple algorithm to generate guesses from a basic set of atoms and operations.  We implemented this algorithm in \textsc{Mathematica} and used it to find new force-free solutions, one of which could represent a rotating force-free vortex near a magnetic null point.  We anticipate that it will be possible to find many more solutions by improving the algorithm and its implementation, experimenting with the choice of primitives, and running for a longer time on faster computers.  While finding solutions is one goal of this approach, we also hope that the reformulation will lead to new insight into the structure of the equations.  We therefore take care to elucidate the mathematical properties of our approach.  We follow the conventions of Ref.~\cite{Gralla2014}.

\section{Force-free magnetic fields in flat spacetime}
\label{sec:ForceFreeMagneticFields}

Stationary force-free configurations with vanishing electric field in flat spacetime are called force-free magnetic fields.  In vector notation, they obey the following equations:
\begin{align}
    \vec{B}=\vec{B}(\vec{x}),\qquad
    \vec{\nabla}\cdot\vec{B}=0,\qquad
    \vec{\nabla}\times\vec{B}=\vec{J},\qquad
    \vec{J}\times\vec{B}=0.
\end{align}
Such fields are also known as ``Beltrami flows'' and form steady solutions of the incompressible Euler equations.\footnote{By the identity $\vec{u}\cdot\vec{\nabla}\vec{u}=\pa{\vec\nabla\times\vec{u}}\times\vec{u}+\vec\nabla\pa{\frac{1}{2}|\vec{u}|^2}$, the field $\vec{u}=\vec{B}$ solves the three-dimensional incompressible Euler equations $\vec{\nabla}\cdot\vec{u}=0$ and $\vec{u}\cdot\vec{\nabla}\vec{u}=-\vec{\nabla}p$ with pressure $p=-\frac{1}{2}|\vec{B}|^2$.  There also exists a different relationship to two-dimensional Eulerian flows.  See e.g., Ref.~\cite{Amari2009} for further discussion.}  In the following, we will consider axisymmetric force-free magnetic fields, 
\begin{align}
\label{eq:Assumptions}
	\L_{\pd_t}F=\L_{\pd_\phi}F=0,\qquad\pd_t\cdot F=0.
\end{align}
Under these assumptions, a degenerate, closed two-form may always be written in the form (see e.g., Ref.~\cite{Gralla2014})
\begin{align}
\label{eq:SimpleForm}
	F=\frac{I}{2\pi\rho}\ed z\wedge\ed\rho+\ed\psi\wedge\ed\phi,
\end{align}
where we work in cylindrical coordinates $\cu{t,\phi,z,\rho}$.  The scalars $I(\rho,z)$ and $2\pi\psi(\rho,z)$ are respectively equal to the electric current and magnetic flux in the upward $z$ direction through a loop of revolution at fixed $(\rho,z)$.\footnote{The scalar $\psi$ must be constant on the rotation axis for the field $F$ to be smooth.  We use the gauge freedom $\psi\to\psi+\textrm{const}$ to make $\psi$ vanish on the rotation axis, in which case it gains the interpretation of the magnetic flux.}  We will refer to these as the \textit{polar current} and \textit{flux function}, respectively.  The magnetic field $B^i=\pa{\star F}^{ti}$ is given by 
\begin{align}
\label{eq:SimpleMagneticField}
	\vec{B}=\frac{\vec{\nabla}\psi\times\hat{\phi}}{\rho}+\frac{I}{2\pi\rho}\hat{\phi},
\end{align}
where $\hat{\phi}=\rho^{-1}\pd_\phi$ and we used the standard orientation $\epsilon_{\rho\phi z}=+\rho$ to define the cross product.  The first term is the poloidal field and the second term is the toroidal (azimuthal) field.  The level sets of $\psi$ are the projections of field lines onto the poloidal plane, or \textit{poloidal field lines}.  These field lines provide the plane foliation that will become our fundamental variable in this paper.

Under the assumptions \eqref{eq:Assumptions} [or equivalently, given the form \eqref{eq:SimpleForm}], the force-free condition implies
\begin{align}
\label{eq:ForceFreeCondition}
	\ed I\wedge\ed\psi=0\qquad\br{\textrm{or equivalently, }I=I(\psi)},
\end{align}
as well as 
\begin{align}
\label{eq:StreamEquation}
	\pd_\rho^2\psi+\pd_z^2\psi-\frac{1}{\rho}\pd_\rho\psi=-\frac{I(\psi)I'(\psi)}{4\pi^2}.
\end{align}
This last equation is called the stream equation.  If $I(\psi)$ is specified as some definite function, then the stream equation is a second-order elliptic partial differential equation.  We may eliminate $I$ from the equation by acting with $\pd_z\psi\pd_\rho-\pd_\rho\psi\pd_z$ on both sides, resulting in
\begin{align}
\label{eq:StreamEquationDerivative}
	\pa{\!\pd_z\psi\pd_\rho-\pd_\rho\psi\pd_z}\pa{\!\pd_\rho^2\psi+\pd_z^2\psi-\rho^{-1}\pd_\rho\psi}=0.
\end{align}
This replaces the pair of Eqs.~\eqref{eq:ForceFreeCondition} and \eqref{eq:StreamEquation} with a single higher-order equation.  Once a solution is found, the current can be reconstructed by integrating Eq.~\eqref{eq:StreamEquation},
\begin{align}
\label{eq:CurrentFromFlux}
	I=\pm\sqrt{-8\pi^2\int\!\ed\psi\,\pa{\pd_\rho^2\psi+\pd_z^2\psi-\frac{1}{\rho}\pd_\rho\psi}}.
\end{align}
If $\psi$ is not a convenient integration variable, one may substitute $\ed\psi=\!\pd_\rho\psi\ed\rho$, $\ed\psi=\!\pd_z\psi\ed z$, or some other convenient choice over suitable domains of the integral.  The integration constant may always be chosen so that the quantity in the square root is positive on any particular region of space where a solution is desired.  The choice of $\pm$ corresponds to the direction of current flow, and its presence follows from the underlying time-reversal invariance of the equations.

\subsection{Foliation approach}

As noted above, the level sets of $\psi$ correspond to the poloidal field lines, which foliate the poloidal plane.  We wish to pass from the field to the foliation as the fundamental variable.  We may describe a foliation as an equivalence class of functions $u(\rho,z)$ whose gradients are parallel and nonvanishing.  That is, two functions $u_1$ and $u_2$ are equivalent if $\pd_\rho u_1=\alpha\pd_\rho u_2$ and $\pd_z u_1=\alpha\pd_z u_2$ for some nonnegative (or nonpositive) function $\alpha(\rho,z)$, or equivalently, if $u_1=f(u_2)$ for an invertible function $f$.  A good equation on foliations $u$ should always be covariant under this gauge freedom $u\to f(u)$.

The stream equation is not a good equation on foliations, since it only holds for a particular representative $\psi$ (namely, the physical magnetic flux).  To pass to an equation on foliations, we let $\psi=\psi(u)$ and eliminate $\psi$ in favor of $u$.  From Eq.~\eqref{eq:ForceFreeCondition}, we then have $I=I(u)$ as well, and hence Eq.~\eqref{eq:StreamEquation} becomes
\begin{align}
\label{eq:StreamEquationFromFoliation}
	A\psi'(u)+B\psi''(u)=-\frac{I(u)I'(u)}{4\pi^2\psi'(u)},
\end{align}
where $A$ and $B$ are given by
\begin{align}
	A=u_{\rho\rho}+u_{zz}-\rho^{-1}u_\rho,\qquad B=u_\rho^2+u_z^2.
\end{align}
(Here and henceforth, we use a subscript to denote partial differentiation.)  To eliminate $\psi$, we take derivatives tangent to the foliation, as done to produce Eq.~\eqref{eq:StreamEquationDerivative}.  For these purposes, we introduce the differential operator (or tangent vector field\footnote{We adopt the viewpoint/definition that vectors are partial differential operators (see e.g., Ref.~\cite{Wald1984}).  Equivalently, the vector $T$ is defined by having $(\rho,z)$ components $(u_z,-u_\rho)$.})
\begin{align}
\label{eq:TangentVector}
	T=u_z\pd_\rho-u_\rho\pd_z.
\end{align}
We will denote the application of $T$ by $\L_T$ (the Lie derivative).  Acting on Eq.~\eqref{eq:StreamEquationFromFoliation} one and two times yields, respectively,
\begin{align}
\label{eq:FoliationEquation}
	\psi'\L_TA+\psi''\L_TB=0,\qquad\psi'\L^2_TA+\psi''\L^2_TB=0,
\end{align}
which can be rewritten as the system 
\begin{align}
\label{eq:FoliationDeterminant}
	\begin{bmatrix}
		\L_TA & \L_TB \\
		\L_T^2A & \L_T^2B
	\end{bmatrix}
	\begin{bmatrix}
		\psi' \\
		\psi''
	\end{bmatrix}
	=
	\begin{bmatrix}
	    0 \\
	    0
	\end{bmatrix}.
\end{align}
The foliation condition for a (nontrivial) solution is simply the vanishing of a determinant,
\begin{align}
\label{eq:FoliationCondition}
	\det
	\begin{bmatrix}
		\L_TA & \L_TB \\
		\L_T^2A & \L_T^2B
	\end{bmatrix}
	=0.
\end{align}
We have now obtained an equation for the foliation representative $u$ without reference to the magnetic flux function $\psi$.  We refer to Eq.~\eqref{eq:FoliationCondition} as the ``foliation condition.''  As shown explicitly below, the field $\psi$ can be reconstructed from any solution satisfying $\L_TB\neq0$ or $\L_TA=\L_TB=0$.

\subsection{Field reconstruction}
\label{sec:FieldReconstruction}

Provided that $\L_TB\neq0$, Eq.~\eqref{eq:FoliationEquation} can be written equivalently as
\begin{align}
\label{eq:FoliationEquationBis}
	\frac{\L_TA}{\L_TB}=-\frac{\psi''}{\psi'}.
\end{align}
The foliation condition \eqref{eq:FoliationCondition} is equivalent to
\begin{align}
\label{eq:FoliationConditionBis}
	\L_T\pa{\frac{\L_TA}{\L_TB}}=0.
\end{align}
This condition ensures that the left-hand side of Eq.~\eqref{eq:FoliationEquationBis} depends only on $u$, so that we may integrate to find
\begin{align}
\label{eq:MagneticFluxReconstruction}
	\psi'(u)=\exp\!\br{-\int\!\ed u\,\frac{\L_TA}{\L_TB}}.
\end{align}
Performing a second integration to obtain $\psi(u)$ is usually not necessary, since only $\ed\psi$ appears in the field strength \eqref{eq:SimpleForm}.  The current $I$ may be reconstructed via Eq.~\eqref{eq:CurrentFromFlux}, or alternatively, from Eq.~\eqref{eq:StreamEquationFromFoliation} by
\begin{align}
\label{eq:CurrentReconstruction}
	I=\pm\sqrt{-8\pi^2\int\!\ed u\,\br{A\pa{\psi'}^2+B\psi''\psi'}},
\end{align}
where again only $\psi'(u)$ appears.  Although $A$ and $B$ are not functions of $u$ alone, the foliation condition guarantees that the integrands in Eqs.~\eqref{eq:MagneticFluxReconstruction} and \eqref{eq:CurrentReconstruction} will only depend on $u$.

If $\L_TB=0$, then Eq.~\eqref{eq:FoliationEquation} requires $\L_TA=0$ as well, so that both $A$ and $B$ are functions of $u$. Equation~\eqref{eq:StreamEquationFromFoliation} then becomes
\begin{align}
\label{eq:LinearStreamEquation}
	2A(u)\br{\psi'(u)}^2+B(u)\frac{d}{du}\cu{\br{\psi'(u)}^2}=-\frac{1}{4\pi^2}\frac{d}{du}\cu{\br{I(u)}^2}.
\end{align}
This is a linear equation in $\br{\psi'(u)}^2$.  Given any current $I(u)$ along the field lines, we can straightforwardly solve for the magnetic flux $\psi(u)$.  A class of solutions to this equation corresponding to the vertical foliation $u=\rho^2$, which obeys $\L_TB=0$, was described in Refs.~\cite{Lundquist1951,Marsh1990}.  The magnetic field admits an arbitrary toroidal component, $B_\phi=\frac{I(u)}{2\pi\rho}$, no radial component, $B_\rho=0$, and has a vertical component $B_z=\partial_\rho\psi$ deduced from \eqref{eq:LinearStreamEquation}.

\subsection{Regularity}

A magnetic field of the form \eqref{eq:SimpleForm} [or equivalently, of the form \eqref{eq:SimpleMagneticField}] is not regular on the axis unless $\psi$ and $I$ both vanish there.\footnote{More generally, $\psi$ may take a constant value on the axis, but one may always shift $\psi$ by this constant without affecting the field strength.  Moreover, only the choice $\psi=0$ on the axis is consistent with the interpretation of $\psi$ as the magnetic flux.}  If $\psi$ does not vanish, then field lines originate from the axis, indicating the presence of a line current of magnetic monopoles.  If $\psi$ does vanish but $I$ does not, then an ordinary electric current flows along the axis.  We may always ensure the vanishing of $\psi$ (lack of magnetic monopoles) by choosing a foliation representative that is constant on the axis.  Such foliations have a field line along the axis.  On the other hand, the vanishing of $I$ (lack of line current) cannot be imposed in all cases, because this demand picks out a unique integration constant in Eq.~\eqref{eq:CurrentReconstruction} [or equivalently, in Eq.~\eqref{eq:CurrentFromFlux}], which may be incompatible with the requirement that the quantity under the square root be positive.

To summarize, force-free solutions satisfying the conditions \eqref{eq:Assumptions} may be constructed by finding solutions to Eq.~\eqref{eq:FoliationCondition} satisfying $u(z=0)=0$ as well as $\L_TB\neq0$, and then using Eqs.~\eqref{eq:SimpleForm}, \eqref{eq:MagneticFluxReconstruction}, and \eqref{eq:CurrentReconstruction} to reconstruct the field strength.  Depending on the foliation, a line current may be required to flow on the axis to support the solution.

\subsection{Function builder and solutions}

One advantage of the foliation equation over the original stream equation is that it makes it far simpler to guess solutions.  The reason is that for each exact solution $\psi$ of the stream equation, there exist an infinite number of solutions $u(\psi)$ to the foliation equation.  One merely needs to chance upon a single representative $u(\rho,z)$ in order to find the exact solution $\psi(\rho,z)$.

To search for solutions, we have designed and implemented a simple algorithm to build representatives $u(\rho,z)$ from basic elements and operations.  We initiate the algorithm at depth 1 with the four building functions $\rho$, $z$, $\rho^2+z^2$, and $\rho/z$.  New functions are built at depth $n+1$ from binary operations among the functions at depth $n-p$ and $p$ with $1\leq p\leq n$.  (Unary operations would only create dependent functions.)  The binary operations that we considered are addition, subtraction, multiplication, division, geometric sum, as well as the operations $(x,y)\to\sqrt{(x-1)^2+y^2}$, $\sqrt{(x+1)^2+y^2}$, $xe^y$, and $x\log y$.  After building the list, we check each function individually, first for regularity and then (if regular) for satisfaction of the foliation constraint.  In practice, we save computational time by only evaluating the foliation constraint at one particular point, which we selected to be $\rho=\frac{4}{5}$, $z=\frac{6}{7}$, using exact arithmetic.  If the constraint is exactly 0 at that point, the constraint is tested in the entire plane.  The resulting solutions are then checked for mutual independence and a list of independent regular solutions is produced.  We perform this step last since its complexity is quadratic in the number of functions, as compared with the linearity of the previous steps.

This algorithm generates approximately 66,000 functions up to depth 4. After imposing regularity and the foliation constraint as well as removing redundancy, we are left with only seven mutually independent foliation representatives.  Of these, six turn out to be vacuum solutions ($I=0$, or more generally, $I=\textrm{const}$).

After selecting the simplest representative $u$, the list of vacuum solutions reads as follows:
\begin{subequations}
\label{eq:NonRotatingSolutions}
\begin{align}
	&\textrm{vertical field (external dipole):}&
	u&=\rho^2=r^2\cos^2{\theta},\\
	&\textrm{X-point (external quadrupole):}&
	u&=\rho^2z=r^3\cos^2{\theta}\sin{\theta},\\
	&\textrm{radial:}&
	u&=1-z/\sqrt{z^2+\rho^2}=1-\cos{\theta},\\
	&\textrm{dipolar:}&
	u&=\rho^2/\pa{z^2+\rho^2}^{3/2}=\sin^2{\theta}/r,\\
	&\textrm{parabolic:}&
	u&=\sqrt{z^2+\rho^2}-z=r(1-\cos{\theta}),\\
	&\textrm{hyperbolic:}&
	u&=\frac{\sqrt{z^2+(\rho-b)^2}-\sqrt{z^2+(\rho+b)^2}}{2b}. 
\end{align}
\end{subequations}
Here, $b$ is a constant which was found by the algorithm to be 1 but which we subsequently generalized to be arbitrary.  Note that we can shift any solution by $z\to z+c$ with constant $c$ and still have a solution.  These vacuum solutions are all known.  The first four arise as multipolar solutions when the equation is separated using spherical coordinates, while the latter two are associated with separation in other coordinate systems.  In the first five cases, the flux function is given by $\psi=\psi_0u$, while in the last case, it is given by $\psi/\psi_0=1-\sqrt{1-u^2}$.  The solutions in this list are vacuum, but all have rotating counterparts that are nonvacuum (see next section), some of which are new.

\begin{figure*}
\centering
\subfigure[\ vertical]{
\includegraphics[width=.25\textwidth]{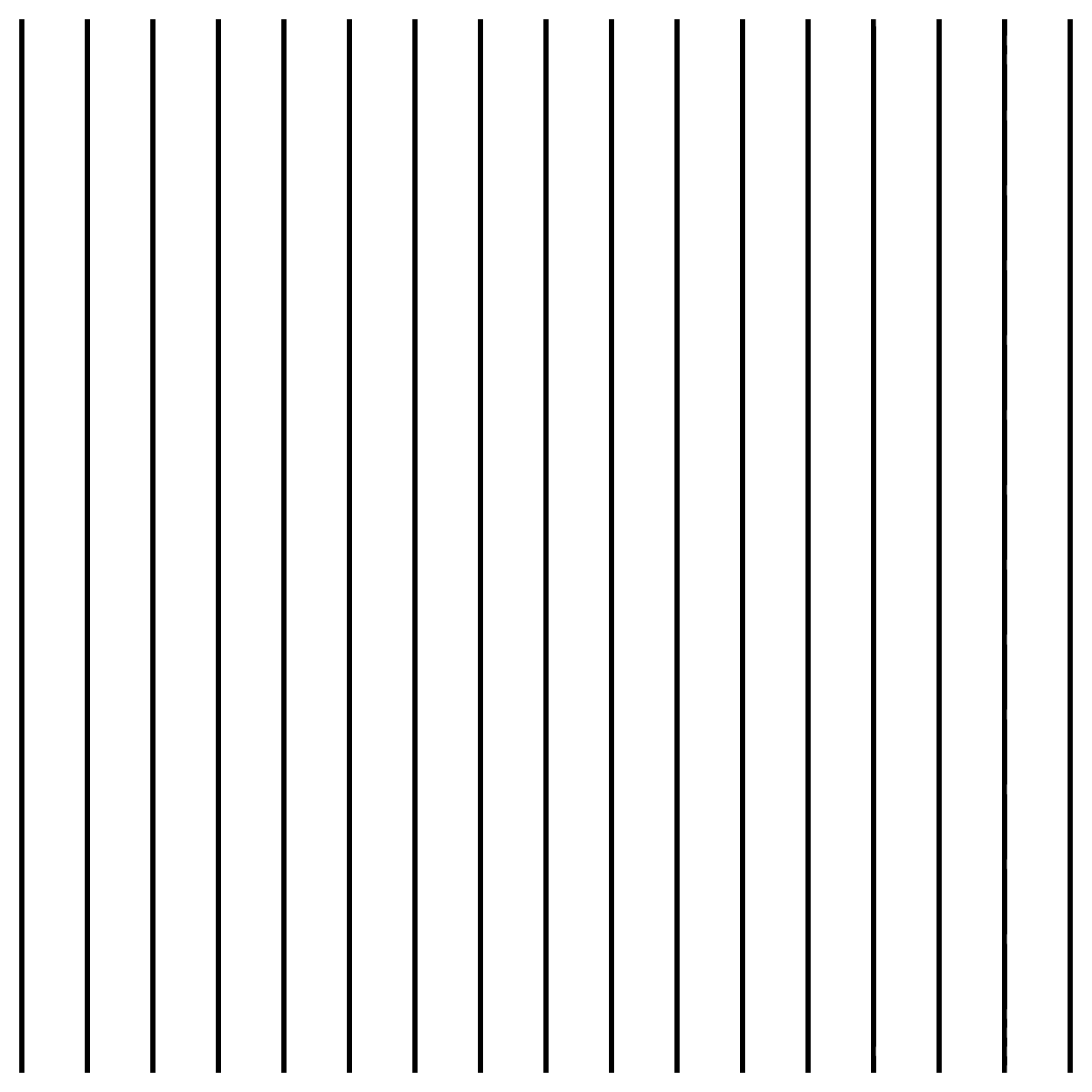}}
\qquad
\subfigure[\ X-point]{
\includegraphics[width=.25\textwidth]{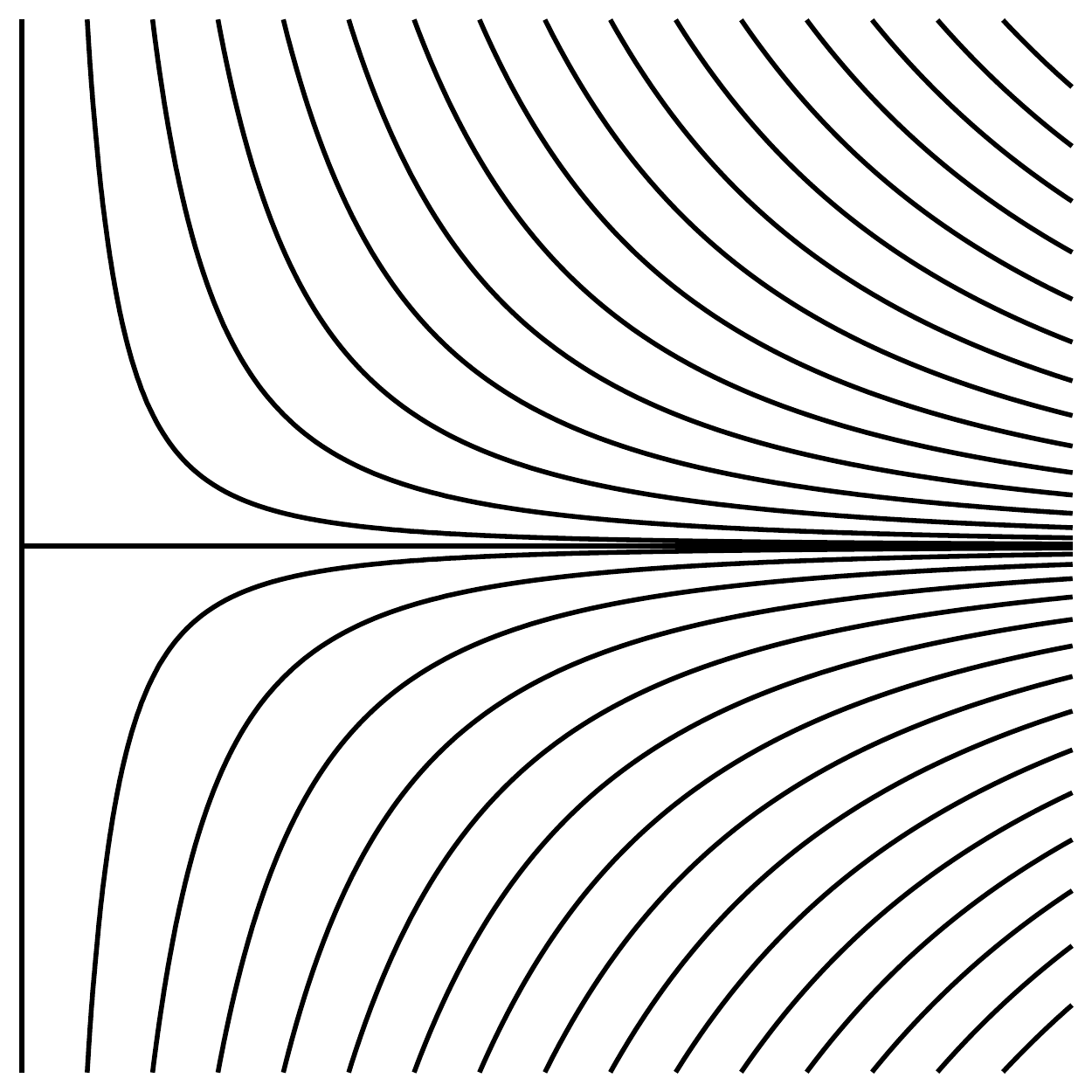}}
\qquad
\subfigure[\ radial]{
\includegraphics[width=.25\textwidth]{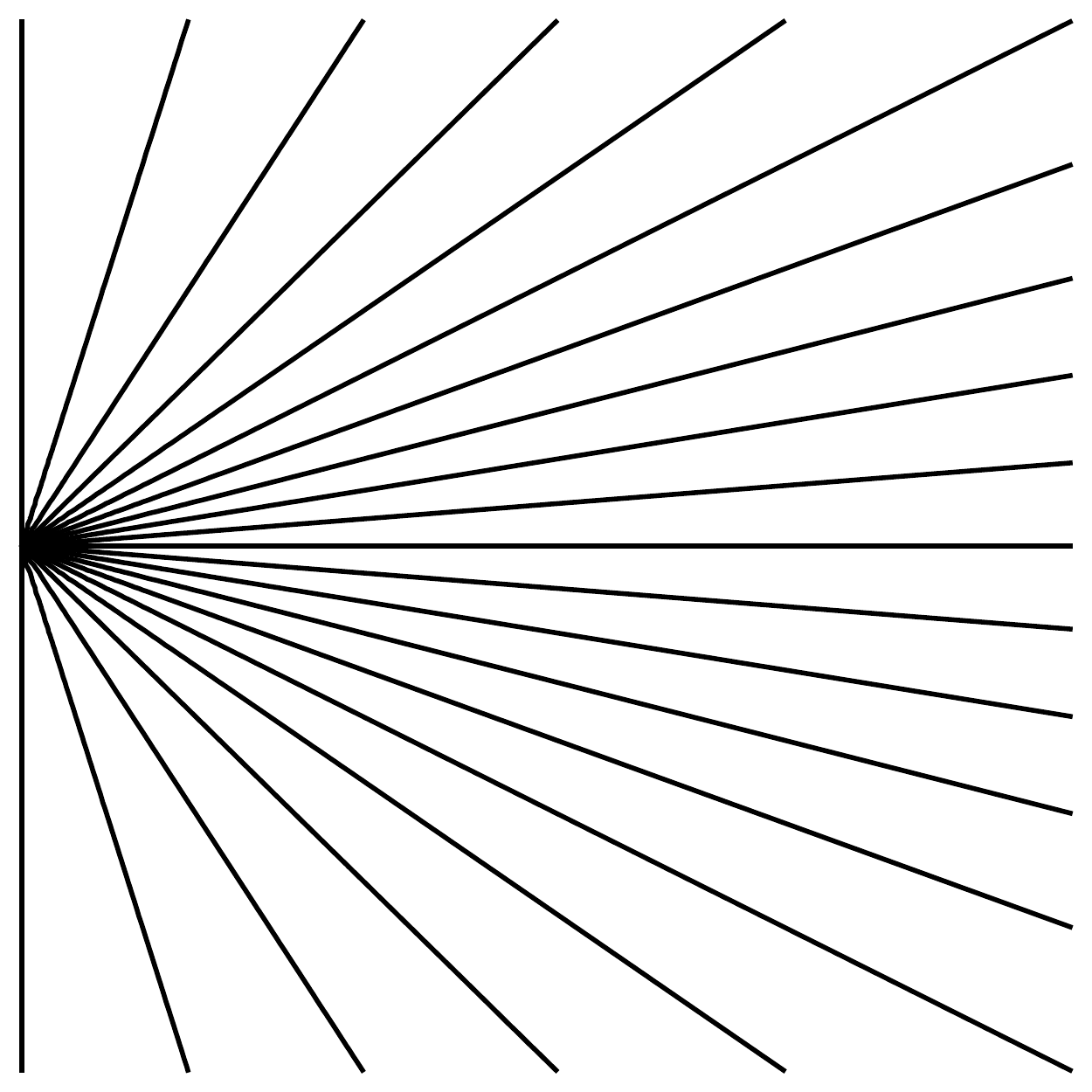}}\\
\subfigure[\ dipolar]{
\includegraphics[width=.25\textwidth]{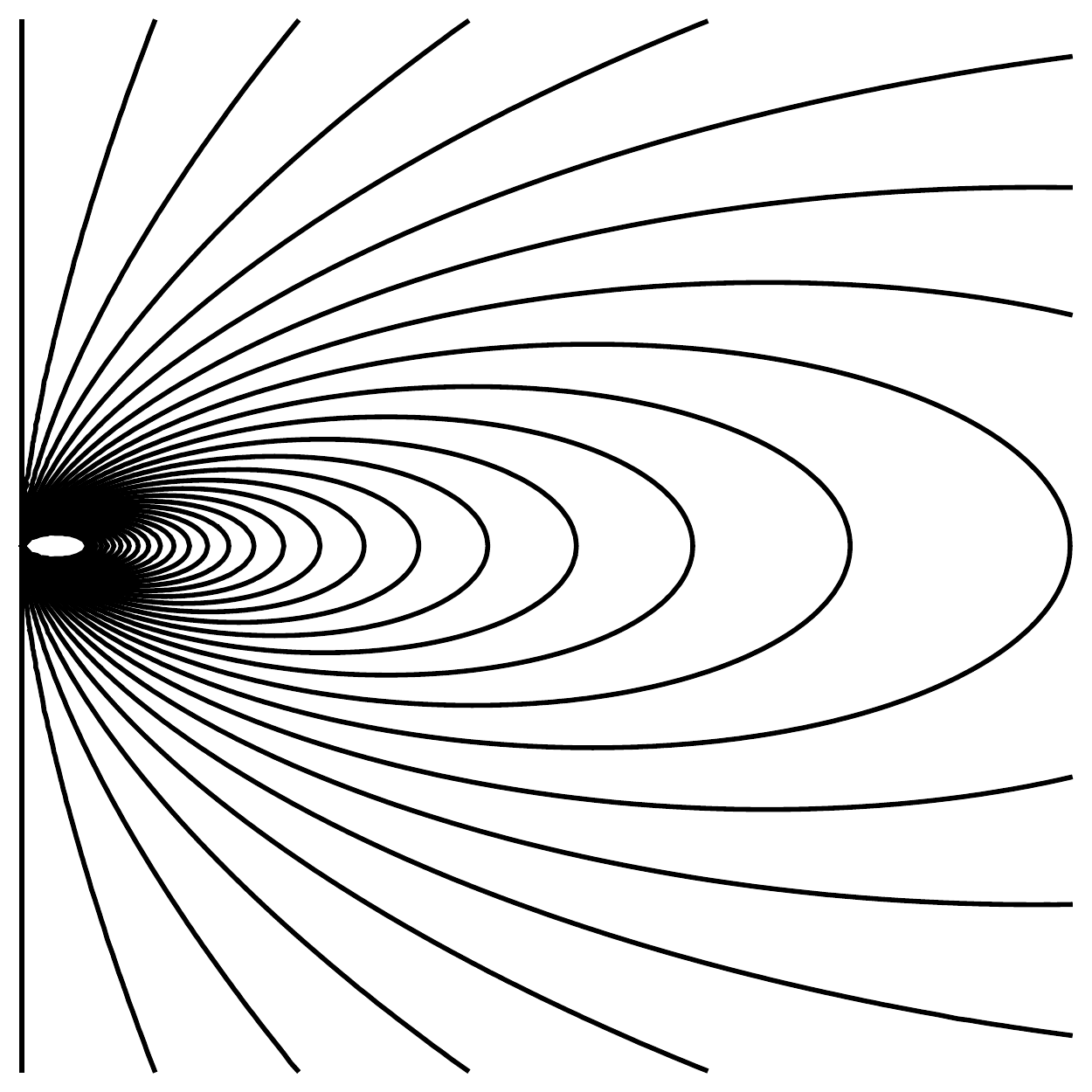}}
\qquad
\subfigure[\ parabolic]{
\includegraphics[width=.25\textwidth]{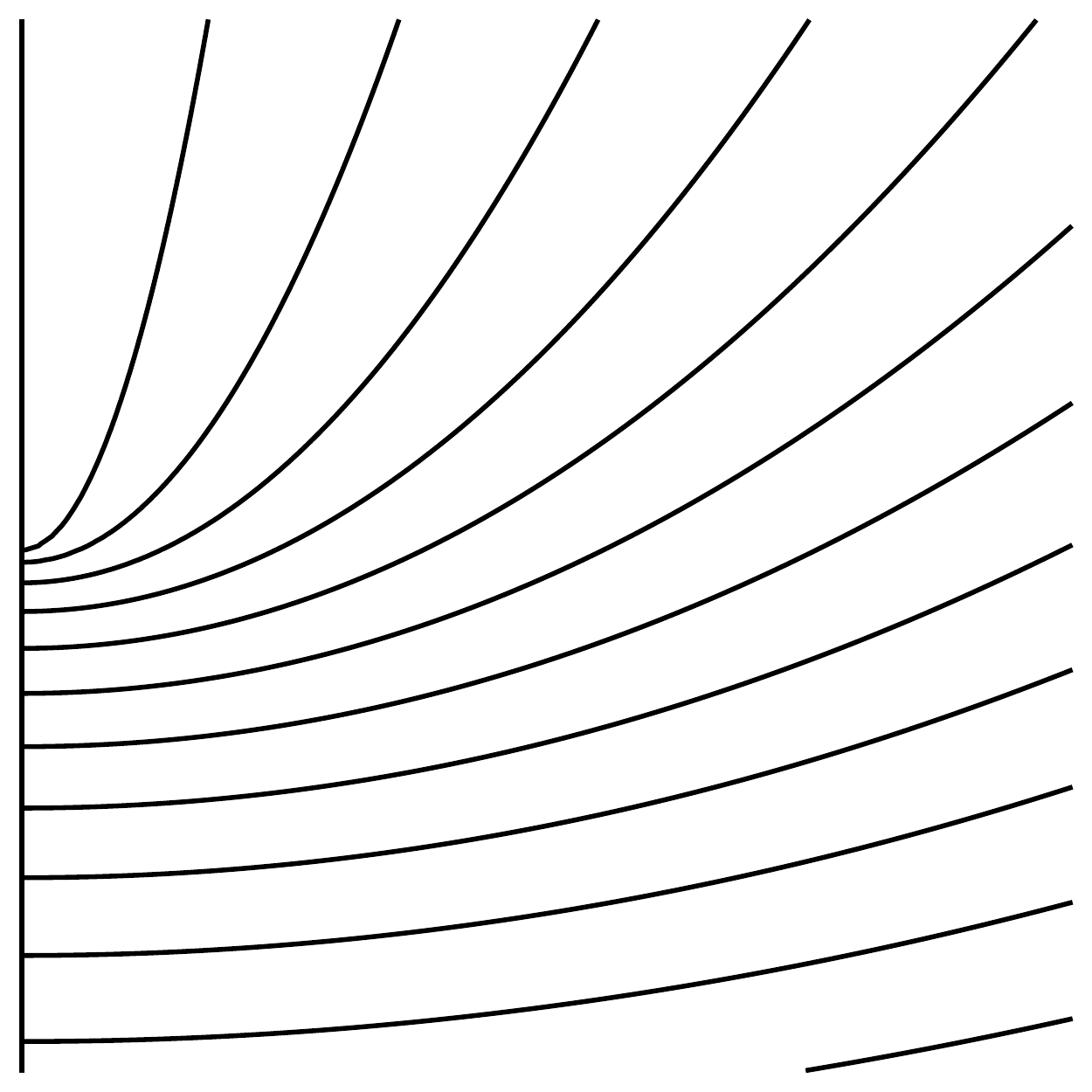}}
\qquad
\subfigure[\ hyperbolic]{
\includegraphics[width=.25\textwidth]{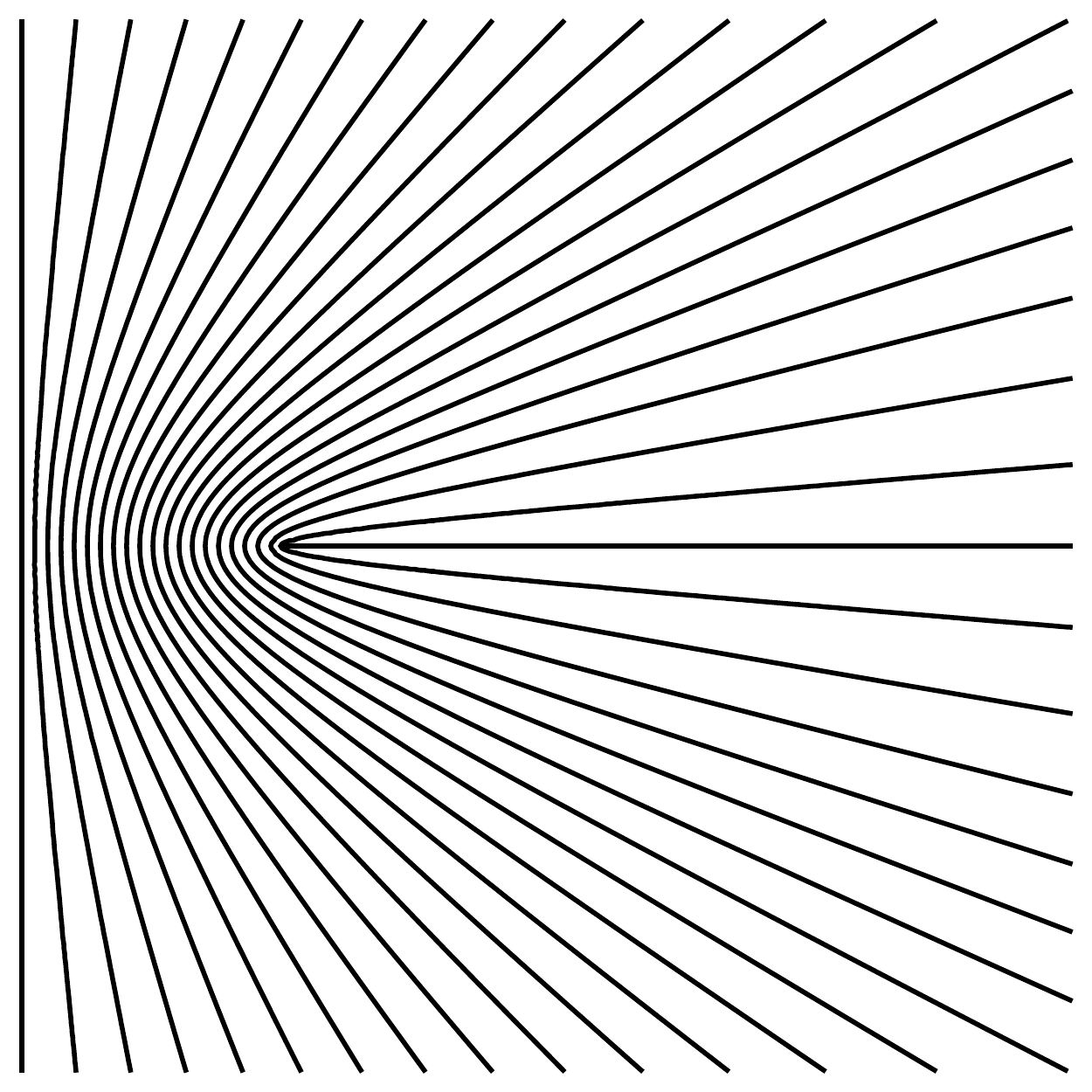}}\\
\subfigure[\ bent]{
\includegraphics[width=.25\textwidth]{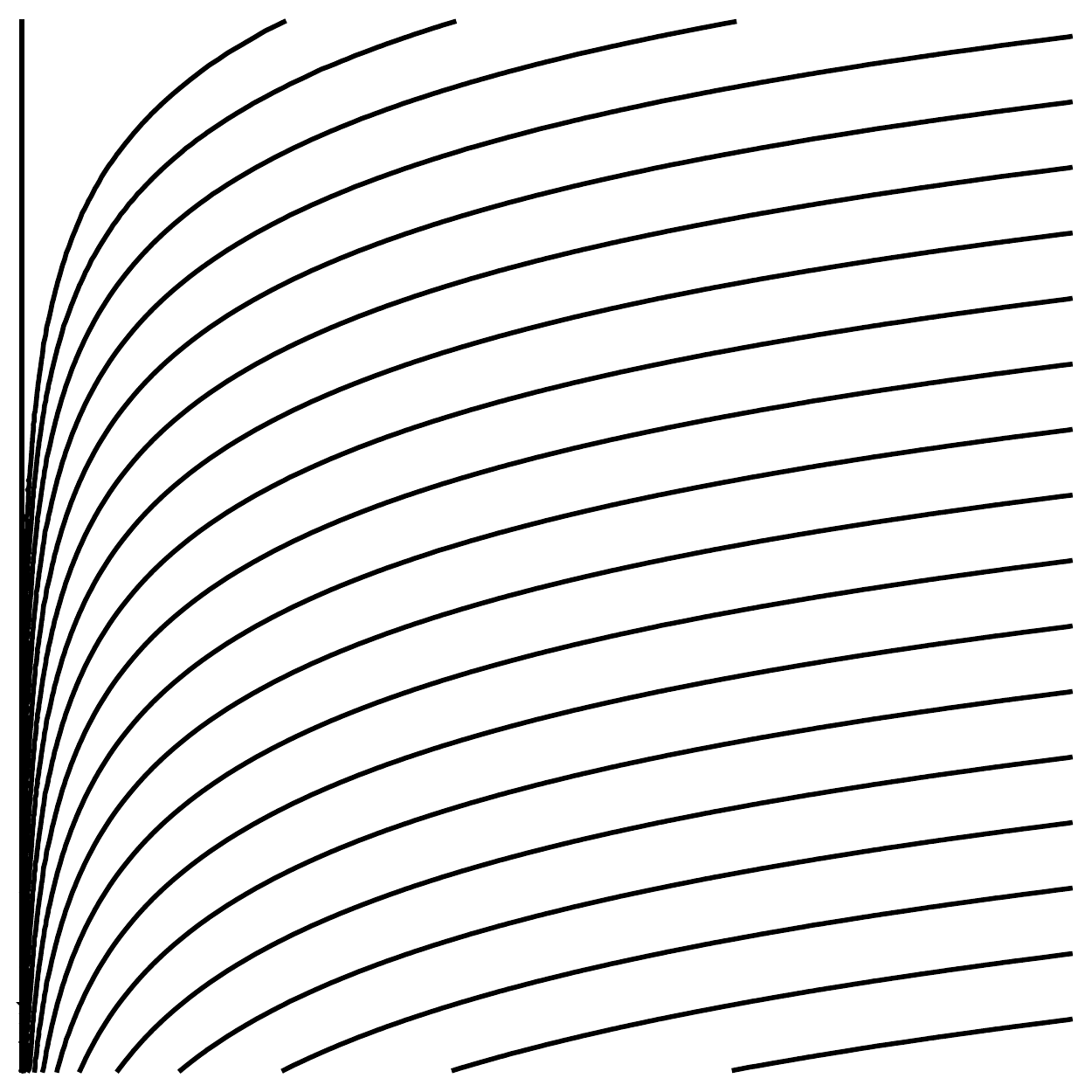}}
\caption{Poloidal field lines of force-free solutions found by the foliation-searching algorithm.  The X-point, dipolar, and bent solutions appear to be new (in the first two cases, when rotation of the field lines is considered).}
\label{fig:foliations}
\end{figure*}

The algorithm finds a single nonvacuum regular solution family,
\begin{align}
\label{eq:BentSolution}
	\textrm{bent:}\qquad
	u=\rho^2e^{-2kz},\qquad\psi=\psi_0u,\qquad I=\pm\sqrt{I_0^2-\pa{4\pi k\psi_0u}^2},
\end{align}
where $k$, $\psi_0$ and $I_0$ are constants.  The field lines are vertical when $k=0$ and bend over for nonzero $k$ (hence the name, ``bent'').  As far as the authors are aware, this solution is new.  In Cartesian coordinates $(x,y,z)$, the magnetic field has components
\begin{align}
	\vec{B}=2\psi_0e^{-2kz}\pa{kx,ky,1}\mp\sqrt{\br{\frac{I_0}{2\pi\pa{x^2+y^2}}}^2-\pa{2k\psi_0e^{-2kz}}^2}\pa{y,-x,0}.
\end{align}
The foliations corresponding to the seven solutions are illustrated in Fig.~\ref{fig:foliations}.

\subsection{Gauge covariance}

An equation for $u(\rho,z)$ can only be considered as an equation for foliations if it holds for all representatives $f(u)$ of the foliation.  In particular, the equation should transform covariantly (i.e., retain its form) under $u\to f(u)$.  This is guaranteed by construction in the derivation of the foliation condition, but it is instructive to check it explicitly.  By direct calculation, the transformation laws for the various quantities are
\begin{subequations}
\label{eq:Covariance}
\begin{align}
	T&\to f'(u)T,\\
	A&\to f'(u)A+f''(u)B,\\
	B&\to\br{f'(u)}^2B,\\
	\L_TA&\to f'(u)\br{f'(u)\L_TA+f''(u)\L_TB},\\
	\L_TB&\to\br{f'(u)}^3\L_TB,\\
	\L_T^2A&\to \br{f'(u)}^2 \br{f'(u)\L^2_TA+f''(u)\L^2_TB},\\
	\L_T^2B&\to\br{f'(u)}^4\L^2_TB.
\end{align}
\end{subequations}
Since the left-hand side of Eq.~\eqref{eq:FoliationCondition} transforms covariantly with an overall factor of $\br{f'(u)}^6$, the invariance of the foliation condition follows.

Note that in the generic case $\L_TB\neq0$, one may always find a gauge where $\L_TA=0$ and $u=\psi$, which amounts to returning to the more basic formulation \eqref{eq:StreamEquationDerivative}.  To do so, one solves $f'(u)\L_TA+f''(u)\L_TB=0$ for $f(u)$.  In particular, one can divide by $\L_TBf'(u)$ and integrate along $u$ since $\L_T\pa{\L_TA/\L_TB}=0$ by Eq.~\eqref{eq:FoliationConditionBis}.

\section{Rotating magnetospheres}

Thus far, we have restricted to vanishing electric field.  When electric fields are included, the general form of a stationary, axisymmetric, degenerate two-form becomes\footnote{More precisely, we consider the conditions \eqref{eq:Assumptions} without $F\cdot\pd_t=0$, but also with $F\cdot\pd_\phi\neq0$ to ensure there is some poloidal field.  An analogous form exists for the case $F\cdot\pd_\phi=0$ \cite{Gralla2014}, and there is an analogous foliation equation which for simplicity, we do not consider here.} [compare to Eq.~\eqref{eq:SimpleForm}]
\begin{align}
	F=\frac{I}{2\pi\rho}\ed z\wedge\ed\rho+\ed\psi\wedge\pa{\!\ed\phi-\Omega\ed t},
\end{align}
for some function $\Omega=\Omega(\psi)$.  The magnetic field sheets discussed in the Introduction are generated by the poloidal field lines $\psi=\textrm{constant}$ rotating with angular velocity $\Omega(\psi)$; hence, $\Omega(\psi)$ is interpreted as the rotation frequency of the field line $\psi$.

The stream equation for rotating magnetospheres reads
\begin{align}
	\br{1-\rho^2\Omega^2(\psi)}\nabla^2\psi-\frac{2}{\rho}\pd_\rho\psi-\rho^2\Omega(\psi)\Omega'(\psi)\pa{\nabla\psi}^2=-\frac{I(\psi)I'(\psi)}{4\pi^2}.
\end{align}
Upon setting $\Omega=0$ and after using the three-dimensional Laplacian, we recover Eq.~\eqref{eq:StreamEquation}.  In the special case of constant $\Omega=\Omega_0$ (``rigid rotation''), the term involving $\Omega'$ does not appear, and one can straightforwardly follow the steps of the previous section to derive a foliation condition.  This condition is again the determinant \eqref{eq:FoliationCondition}, except with the following new definitions for $A$ and $B$:
\begin{align}
	A&=\pa{1-\rho^2\Omega_0^2}\pa{u_{\rho\rho}+u_{zz}}-\frac{1+\rho^2\Omega_0^2}{\rho}u_\rho,\\
	B&=\pa{1-\rho^2\Omega_0^2}\pa{u_\rho^2+u_z^2}.
\end{align}
One can search for rigidly rotating magnetospheres by running the algorithm we described in the previous section with this new choice of $A$ and $B$. 

In the general case $\Omega=\Omega(\psi)$, it is not possible to eliminate $\Omega$ from the equation by the strategy we have pursued.  Instead, applying the same manipulations as above, we again obtain Eq.~\eqref{eq:StreamEquationFromFoliation}, but this time with 
\begin{align}
\label{eq:A}
	A&=\br{1-\rho^2\Omega^2(u)}\pa{u_{\rho\rho}+u_{zz}}-\frac{1+\rho^2\Omega^2(u)}{\rho}u_\rho-\rho^2\Omega(u)\Omega'(u)\pa{u_\rho^2+u_z^2},\\
\label{eq:B}
	B&=\br{1-\rho^2\Omega^2(u)}\pa{u_\rho^2+u_z^2}.
\end{align}
Equation~\eqref{eq:FoliationCondition} is now a first-order differential equation for $\Omega(u)$ which is consistent only for specific foliations where all fields are $u$ dependent only.  If consistent, it can then be solved for $\Omega(u)$, and the current $I(u)$ can then be reconstructed from Eq.~\eqref{eq:CurrentReconstruction}.  

While it is difficult to directly solve the foliation condition \eqref{eq:StreamEquationFromFoliation} with $A$ and $B$ as given in Eqs.~\eqref{eq:A} and \eqref{eq:B}, it is straightforward to check by hand whether the nonrotating solutions \eqref{eq:NonRotatingSolutions} have rotating counterparts.  One must simply evaluate Eq.~\eqref{eq:FoliationCondition} assuming one of the seven foliations found in Sec.~\ref{sec:ForceFreeMagneticFields}, and then check consistency.  We find that \textit{all} of the nonrotating solutions found in Sec.~\ref{sec:ForceFreeMagneticFields} have rotating counterparts.\footnote{However, superpositions of the vacuum solutions \eqref{eq:NonRotatingSolutions}, which are valid when $\Omega=0$, do not admit rotating generalizations.}  Four of them (vertical, radial, parabolic, hyperbolic) are special in that any $\Omega(u)$ gives a solution.  These four solutions were all previously known \cite{Michel1973,Blandford1976,Beskin1992,Gralla2016a}, but we present them again here for the sake of completeness:
\begin{subequations}
\begin{align}
	&\textrm{vertical:}&
	\psi&=\psi_0u,\quad\Omega=\Omega(u),\quad I=\pm4\pi\psi_0u\Omega(u),\\
	&\textrm{radial:}&
	\psi&=\psi_0u,\quad\Omega=\Omega(u),\quad I=\pm2\pi\psi_0u\pa{2-u}\Omega(u),\\
	&\textrm{parabolic:}&
	\psi&=\psi_0\int\!\frac{\ed u}{\sqrt{1+\br{u\Omega(u)}^2}},\quad\Omega=\Omega(u),\quad I=\pm\frac{4\pi\psi_0u\Omega(u)}{\sqrt{1+\br{u\Omega(u)}^2}},\\
	&\textrm{hyperbolic:}&
	\psi&=\psi_0\int\!\frac{u\ed u}{\sqrt{1-u^2}\sqrt{1-\br{b\Omega(u)u^2}^2}},\quad\Omega=\Omega(u),\quad I=\pm\frac{2\pi\psi_0\Omega(u)u^2}{\sqrt{1-\br{b\Omega(u)u^2}^2}}. 
\end{align}
\end{subequations}
We have chosen the integration constant $I_0$ to ensure global regularity.

The remaining three solutions require special choices of $\Omega(u)$ and appear to be new:
\begin{align}
	&\textrm{X-point:}&
	u&=\rho^2z,\quad\psi=\psi_0u,\quad\Omega=\Omega_0,\quad I=\pm4\pi\psi_0\Omega_0\sqrt{I_0^2+u^2},\\
	&\textrm{dipolar:}&
	u&=\frac{\rho^2}{\pa{z^2+\rho^2}^{3/2}},\quad\psi=\psi_0u,\quad\Omega=\frac{\Omega_0}{u^2},\quad I=\pm4\pi\psi_0\Omega_0\sqrt{I_0^2+\frac{1}{u^2}},\\
	&\textrm{bent:}&
	u&=\rho^2e^{-2kz},\quad\psi=\psi_0u,\quad\Omega=\frac{\Omega_0}{u},\quad I=\pm4\pi k\psi_0\sqrt{I_0^2-u^2}.
\end{align}
In these solutions, we include the integration constant $I_0$.  The dipolar and bent solutions have singularities and hence could only be realized over a finite region.  On the other hand, we may ensure global regularity of the X-point solution by fixing the integration constant to be $I_0=0$.  This solution is a rotating quadrupolar field configuration and could represent a ``force-free vortex'' near a magnetic null point.  We have not checked whether the configuration is stable.

\section{Generalization to curved spacetime}

We now generalize to an arbitrary stationary, axisymmetric, circular spacetime.  For the basic formulation of the force-free equations, we follow the approach and notation of Ref.~\cite{Gralla2014}.  We work in coordinates $\cu{t,\phi,x^a}$, where the timelike and axial Killing fields are $\pd_t$ and $\pd_\phi$, respectively, and $x^a$ are two arbitrary poloidal coordinates (such as Boyer-Lindquist $r$ and $\theta$ for the Kerr metric).  The metric takes the general form
\begin{align}
\label{eq:Metric}
	ds^2=-\alpha^2\ed t^2+\rho^2\pa{\!\ed\varphi-\Omega_Z\ed t}^2+g^P_{ab}\ed x^a\ed x^b,
\end{align}
where all metric components depend only on the poloidal coordinates $x^a$.  Here, $\rho$ is the cylindrical radius, while $\alpha$ and $\Omega_Z$ respectively denote the redshift factor and angular frequency of observers orbiting at fixed $\theta$ with zero angular momentum.

A degenerate, stationary, axisymmetric field with $F\cdot\pd_\phi\neq0$ may always be written as
\begin{align}
	F=\frac{I(\psi)}{2\pi\alpha\rho}\epsilon_P+\ed\psi\wedge\br{\!\ed\phi-\Omega(\psi)\ed t},
\end{align}
where $\epsilon_P$ is the metric-compatible poloidal volume element given by $\sqrt{g^P}$ times the Levi-Civit\`a symbol.  Here, $\psi$, $I$, and $\Omega$ have the same physical interpretation as in flat spacetime.  The stream equation is
\begin{align}
	\alpha\rho\,\nabla_a\!\pa{\alpha\rho\ab{\eta}^2\nabla^a\psi}+\rho^2\Omega'(\psi)\br{\Omega(\psi)-\Omega_Z}\nabla_a\psi\nabla^a\psi=-\frac{I(\psi)I'(\psi)}{4\pi^2},
\end{align}
where $\nabla_a$ is the covariant derivative compatible with the poloidal metric, and we have introduced the one-form $\eta\equiv\ed\phi-\Omega(\psi)\ed t$.  In terms of the functions appearing in the metric \eqref{eq:Metric}, we have
\begin{align}
	\ab{\eta}^2=\frac{1}{\rho^2}-\frac{\br{\Omega(\psi)-\Omega_Z}^2}{\alpha^2}.
\end{align}

We now follow the steps of the previous sections to derive the foliation condition.  We consider an arbitrary foliation representative $u$ on the poloidal plane, and assume without loss of generality that $\psi$ is a function of $u$.  We define the length of the gradient $\ell$ as well as the unit normal and tangent vectors,
\begin{align}
\label{eq:FoliationVectors}
	\ell=\sqrt{\nabla^au\nabla_au},\qquad n^a=\nabla^au/\ell,\qquad m^a=\epsilon_P^{ab}n_b.
\end{align}
For the tangent vector $T^a$, one may choose any normalization, which we fix as
\begin{align}
\label{eq:GeometricTangentVector}
	T^a=\ell\sqrt{g^P}m^a.
\end{align}
This choice ensures that $T=u_z\pd_\rho-u_\rho\pd_z$, in agreement with the simple form \eqref{eq:TangentVector} that the vector field took in flat spacetime.  The foliation condition once again takes the determinant form \eqref{eq:FoliationCondition} (or $\L_TA=\L_TB=0$), but with $A$ and $B$ now given by 
\begin{align}
\label{eq:GeneralA}
	A&=\alpha^2\rho^2\ab{\eta}^2\ell\br{\kappa_n+\L_n\log\ab{\alpha\rho\ab{\eta}^2\ell}}+\rho^2\ell^2\Omega'(u)\br{\Omega(u)-\Omega_Z},\\
\label{eq:GeneralB}
	B&=\alpha^2\rho^2\ab{\eta}^2\ell^2,
\end{align}
where $\kappa_n=\nabla_an^a$ is the extrinsic curvature of the foliation.  The flux and current can be reconstructed in the same way as before, using the formulas in Sec.~\ref{sec:FieldReconstruction}.

We have used the foliation condition to rederive two known solutions corresponding to foliations $u=\cos{\theta}$ and $u=r$ in the Kerr metric in Boyer-Lindquist coordinates.  Unfortunately, we did not find any new solutions in the Kerr metric.  The perfectly radial foliation $u=\cos{\theta}$ was found in Ref.~\cite{Menon2007} and satisfies the foliation condition in Kerr provided that we define
\begin{align}
	\Omega(u)=\frac{1}{a\pa{1-u^2}}. 
\end{align}
The orthogonal, or dual, perfectly circular foliation $u=r$ was found in Ref.~\cite{Menon2015}.  It satisfies the foliation condition provided that we define
\begin{align}
	\Omega(u)=\frac{a}{u^2+a^2}. 
\end{align}
However, this foliation is not tangent to the $z$ axis and hence, the associated solution is singular.

\section{Geometric formulation}

We have emphasized that meaningful equations on foliations must transform covariantly under $u\to f(u)$.  The form \eqref{eq:FoliationCondition} does not make covariance manifest, since none of the building blocks \eqref{eq:Covariance} are individually covariant.  This is akin to expressing the Einstein field equations in terms of partial derivatives rather than (spacetime-)covariant derivatives, after which covariance of the entire equation would be seen only after many miraculous cancellations.  It is clearly preferable to have manifestly covariant expressions, which requires expressing all quantities in terms of geometric invariants of the foliation. 

The fundamental building blocks of this geometric formulation are the extrinsic curvatures of the foliation of unit normal $n^a$ and of the orthogonal foliation of unit normal $m^a$,
\begin{align}
	\kappa_n&=\nabla_an^a=g^{ab}\nabla_an_b=m^am^b\nabla_an_b,\\
	\kappa_m&=\nabla_am^a=\epsilon^{ab}\nabla_an_b=-n^am^b\nabla_an_b. 
\end{align}
All nongeometric quantities (such as $\ell$) need to be replaced by objects that are invariant under $u\to f(u)$.  In order to describe all quantities of interest, it is convenient to introduce the additional invariants
\begin{align}
\label{eq:GeometricInvariants}
	\Delta_n=\L_n\log\ab{\alpha \rho \ab{\eta}^2}+\kappa_n,\qquad\Delta_m=\L_m\log\ab{\alpha^2\rho^2\ab{\eta}^2}-\kappa_m,\qquad\alpha_m=\L_m\log\ab{\alpha^2\ab{\eta}^2}.
\end{align}
All factors of $\alpha$, $\rho$, and $|\eta|^2$ can be expressed in terms of these invariants. Finally, $\Omega_Z$ is another independent invariant.  In Appendix \ref{app:Derivation}, we reformulate the foliation condition \eqref{eq:FoliationCondition} in terms of invariants, leading to
\begin{align}
\label{eq:GeometricCondition}
	\det\begin{bmatrix}
		X+X_\Omega && \Delta_m-\kappa_m \\
		\pa{\vec\L_m+\Delta_m}X+\pa{\vec\L_m+\kappa_m}X_\Omega && \L_m\pa{\Delta_m-\kappa_m}
	\end{bmatrix}
	=0,
\end{align}
where
\begin{align}
	X&=\pa{\vec\L_m+\kappa_m}\Delta_n-\pa{\vec\L_n+\kappa_n}\kappa_m=\nabla_a\br{\Delta_nm^a-\kappa_mn^a},\\
	X_\Omega&=\frac{\ell}{\alpha^2 |\eta|^2}\Omega'(u)\cu{\br{\Omega(u)-\Omega_Z}\pa{\Delta_m-\kappa_m-\alpha_m}-\L_m\Omega_Z}.
\end{align}
The notation $\vec\L_m$ emphasizes that the Lie derivative is an operator acting on its argument on the right.  All quantities in Eq.~\eqref{eq:GeometricCondition} are manifestly invariant under reparameterizations of the foliation.  

The condition $\L_TB\neq0$ is equivalent to $\Delta_m-\kappa_m\neq0$.  When $\Omega'(u)=0$, the condition $\L_TB=\L_TA=0$ is equivalent to $X=0$ and $\Delta_m-\kappa_m=0$.  Finally, note that the foliation condition \eqref{eq:GeometricCondition} is homogeneous of degree 4 in derivatives along poloidal coordinates.  Indeed, the diagonal entries of the matrix \eqref{eq:GeometricCondition} are homogeneous of degree 2, the upper right entry is homogeneous of degree 1, and the lower left entry is homogeneous of degree 3. 

\section*{Acknowledgments}

G.C. would like to thank the Center for Mathematical Sciences and Applications as well as the Center for the Fundamental Laws of Nature at Harvard for their hospitality while part of this work was conducted.

G.C. is a Research Associate of the Fonds de la Recherche Scientifique F.R.S.-FNRS (Belgium) and he acknowledges the current support of the ERC Starting Grant No. 335146 ``HoloBHC'' and convention IISN 4.4503.15 of FNRS-Belgium.  S.G. and A.L. were supported by NSF Grant No. 1205550 to Harvard University.  S.G. was also supported by NSF Grant No. 1506027 to the University of Arizona.  Finally, we thank anonymous referees for their comments.

\appendix

\section{The foliation determines the field}
\label{app:Foliation}

Every degenerate, closed two-form $F_{\mu\nu}$ defines a foliation of spacetime into two-dimensional submanifolds spanned by the vectors $v^\mu$ such that $F_{\mu\nu}v^\mu=0$  \cite{Carter1979,Uchida1997,Gralla2014}.  In particular, a force-free solution defines a foliation. The converse is not true, but we now show that in the timelike case $(F_{\mu\nu}F^{\mu\nu}>0)$ of physical interest, \textit{if} a foliation has an associated force-free solution, then that solution is unique.  The result also holds in the spacelike case, but not for null foliations.

We use the Newman-Penrose formulation \cite{Newman1962} and (in this appendix only) work in the signature $(+,-,-,-)$.  The two-form is represented in terms of three complex scalars by
\begin{align}
	\phi_0&=F_{\mu\nu}\ell^\mu m^\nu,\\
	\phi_1&=\frac{1}{2}F_{\mu\nu}\pa{\ell^\mu n^\nu+\bar{m}^\mu m^\nu},\\
	\phi_2&=F_{\mu\nu}\bar{m}^\mu n^\nu,
\end{align}
where the null tetrad $\cu{\ell^\mu,n^\mu,m^\mu,\bar{m}^\mu}$ satisfies $\ell\cdot n=1$ and $m\cdot\bar{m}=-1$, with all other inner products vanishing.  (Here, $\ell$ and $n$ are real null vectors, while $m$ is a complex null vector.)  Given a timelike foliation, we may erect a Newman-Penrose tetrad by taking $\ell$ and $n$ to lie in the foliation.  In particular, $F\cdot\ell=F\cdot n=0$, so we have
\begin{align}
\label{eq:MagneticFieldScalars}
	\phi_0=\phi_2=0,\qquad\phi_1=\frac{iB}{2}.
\end{align}
Here, $B=\sqrt{F_{\mu\nu}F^{\mu\nu}/2}$ is the magnetic field strength.

With the conditions \eqref{eq:MagneticFieldScalars}, Maxwell's equations become (see e.g., Ref~\cite{Teukolsky1973})
\begin{align}
	\pa{\ell\cdot\nabla-2\rho}\phi_1&=2\pi J_\ell,\\
	-\pa{n\cdot\nabla+2\mu}\phi_1&=2\pi J_n,\\
	\pa{m\cdot\nabla-2\tau}\phi_1&=2\pi J_m,\\
	-\pa{\bar{m}\cdot\nabla+2\pi}\phi_1&=2\pi J_{\bar{m}}.
\end{align}
The scalars $\rho,\mu,\tau,\pi$ on the left-hand sides are spin coefficients that characterize derivatives of the tetrad vectors \cite{Newman1962}.  (The $\pi$'s on the right-hand sides are just the usual number $3.14\dots$)  We also introduce the projection of $J$ onto the null tetrad, e.g., $J_\ell=J\cdot\ell$.  The current is reconstructed by $J=J_\ell n+J_n\ell-J_m\bar{m}-J_{\bar{m}}m$.

Since $\ell$ and $n$ span a surface, we have $\bar{\tau}=-\pi$ by Proposition (4.14.3) of Ref.~\cite{Penrose1984}, making the last two equations equivalent.  The condition $F\cdot J=0$ becomes $J_m=J_{\bar{m}}=0$, and the force-free equations are
\begin{subequations}
\label{eq:Transport}
\begin{align}
	\pa{\ell\cdot\nabla-2\textrm{Re}[\rho]}B&=0,\\
	\pa{n\cdot\nabla+2\textrm{Re}[\mu]}B&=0,\\
	\pa{m\cdot\nabla-2\tau}B&=0.
\end{align}
\end{subequations}
The remaining two equations, $2\pi J_l=\textrm{Im}[\rho]B$ and $2\pi J_n=-\textrm{Im}[\mu]B$ serve to compute the current once $B$ is found.  Equations~\eqref{eq:Transport} are four transport equations for the single scalar $B$, which uniquely determine the solution if it exists. 

For most foliations, no consistent solution of Eqs.~\eqref{eq:Transport} will exist.  Determining the integrability conditions in terms of geometric properties of the foliation would constitute the general foliation formulation of force-free electrodynamics.  While integrability conditions for Eqs.~\eqref{eq:Transport} can be determined by working out the commutators of the relevant differential operators, the result is a complicated expression that depends on arbitrary choices in erecting the tetrad in addition to the geometric properties of the foliation.  It would be desirable to eliminate (or at least understand) this gauge arbitrariness to produce what could be called the foliation formulation of force-free electrodynamics.

\section{Detailed derivations}
\label{app:Derivation}

In this appendix, we present the derivation of the geometric form of the foliation condition, Eq.~\eqref{eq:GeometricCondition}.  All calculations are done using the poloidal metric and volume element, as indicated by the continued use of latin indices.

\subsection{Preliminaries}

Recall from Eq.~\eqref{eq:GeometricTangentVector} that we defined $T^a=\ell\sqrt{g^P}m^a$. This implies that
\begin{align}
	\L_T\phi=\ell\sqrt{g^P}\L_m\phi.
\end{align}
The definition of $m^a$ in Eq.~\eqref{eq:FoliationVectors} can be inverted to give the useful relation
\begin{align}
	n^a=-\epsilon_P^{ab}m_b.
\end{align}
It then follows that
\begin{align*}
	\kappa_m=\nabla_am^a
	=\nabla_a\pa{\epsilon_P^{ab}n_b}
	=\nabla_a\pa{\epsilon_P^{ab}\frac{\nabla_bu}{\ell}}
	=\epsilon_P^{ab}\nabla_a\pa{\frac{1}{\ell}}\nabla_bu+\frac{1}{\ell}\epsilon_P^{ab}\nabla_a\nabla_bu,
\end{align*}
where in the last step, we used the compatibility of the Levi-Civit\`a tensor $\epsilon_P$.  Since the Riemann tensor always vanishes in two dimensions, $\nabla_a\nabla_bu$ is symmetric.  Hence, its contraction $\epsilon_P^{ab}\nabla_a\nabla_bu$ with the antisymmetric symbol vanishes, leaving
\begin{align*}
	\kappa_m=\epsilon^{ab}\nabla_a\pa{\frac{1}{\ell}}\nabla_bu
	=-\frac{1}{\ell^2}\epsilon^{ab}\nabla_a\ell\nabla_bu
	=-\frac{1}{\ell}\pa{\epsilon^{ab}\frac{\nabla_bu}{\ell}}\nabla_a\ell
	=-\frac{1}{\ell}m^a\nabla_a\ell
	=-\frac{1}{\ell}\L_m\ell.
\end{align*}
As such, we have established that
\begin{align}
\label{eq:CurvatureLength}
	\L_m\log\ell=-\kappa_m.
\end{align}
For future reference, note also that
\begin{align}
\label{eq:UsefulIdentity}
	n^b\nabla_an_b=\frac{1}{2}\nabla_a\pa{n^bn_b}
	=\frac{1}{2}\nabla_a1
	=0.
\end{align}
Next, we define the acceleration of the foliation,
\begin{align}
	\alpha^a=n^b\nabla_bn^a,
\end{align}
which obeys $n_a\alpha^a=0$, and therefore, $\alpha^a\propto m^a$. As such, there exists some proportionality constant $\lambda$ such that $\alpha^a=\lambda m^a$. From the unit normalization of $m^a$, we see that $m^a\alpha_a=\lambda m^am_a=\lambda$. Hence,
\begin{align*}
	\lambda=m^a\alpha_a
	=m^an^b\nabla_bn_a
	=\pa{m^an^b-n^am^b}\nabla_bn_a
	=\pa{g^{ac}g^{bd}-g^{ad}g^{bc}}m_cn_d\nabla_bn_a,
\end{align*}
where in the penultimate step we used Eq.~\eqref{eq:UsefulIdentity} to see that $n^am^b\nabla_bn_a=0$. Invoking the geometric identity $g^{ac}g^{bd}-g^{ad}g^{bc}=\epsilon^{ab}\epsilon^{cd}$, we find that
\begin{align*}
	\lambda=\epsilon_P^{ab}\epsilon_P^{cd}m_cn_d\nabla_bn_a
	=m_c\pa{\epsilon_P^{cd}n_d}\nabla_b\pa{\epsilon_P^{ab}n_a}
	=m_cm^c\nabla_b\pa{-m^b}
	=-\nabla_bm^b
	=-\kappa_m.
\end{align*}
As such, the acceleration of the foliation is related to its normalized tangent by
\begin{align}
\label{eq:AccelerationTangent}
	\alpha^a=-\kappa_mm^a.
\end{align}
The extrinsic curvature is defined as
\begin{align}
	K_{ab}=\nabla_an_b-n_a\alpha_b.
\end{align}
Note that
\begin{align}
	n^aK_{ab}&=n^a\nabla_an_b-n^an_a\alpha_b=\alpha_b-\alpha_b=0,\\
	n^bK_{ab}&=n^b\nabla_an_b-n^bn_a\alpha_b=0-n^bn_a\pa{-\kappa_mm_b}=0,
\end{align}
where in the second line, we used Eq.~\eqref{eq:UsefulIdentity} together with the orthogonality condition $n^bm_b=0$.  Since the projections of $K_{ab}$ along $n^a$ all vanish, it results that we must necessarily have
\begin{align}
	K_{ab}=\tau m_am_b
\end{align}
for some proportionality constant $\tau$, which may be determined from the unit normalization of $m^a$:
\usetagform{brackets}
\begin{align*}
	\tau&=\tau\pa{m^am_a}\pa{m^bm_b}
	=m^am^bK_{ab}
	=m^am^b\pa{\nabla_an_b-n_a\alpha_b}\\
	&=m^am^b\pa{\nabla_an_b+\kappa_mn_am_b}
	=m^am^b\nabla_an_b
	\tag{by Eq.~(\ref{eq:AccelerationTangent})}\\
	&=m^am^b\nabla_an_b+n^an^b\nabla_an_b=\pa{m^am^b+n^an^b}\nabla_an_b
	\tag{by Eq.~(\ref{eq:UsefulIdentity})}\\
	&=\pa{m^a\epsilon^{bc}n_c-n^a\epsilon^{bc}m_c}\nabla_an_b
	=\pa{m^an_c-n^am_c}\nabla_a\pa{\epsilon^{bc}n_b}\\
	&=-\pa{m^an^c-n^am^c}\nabla_a\pa{\epsilon_{cb}n^b}
	=-\pa{m^an^c-n^am^c}\nabla_am_c\\
	&=-m_bn_d\pa{g^{ab}g^{cd}-g^{ad}g^{bc}}\nabla_am_c
	=-m_bn_d\epsilon^{ac}\epsilon^{bd}\nabla_am_c\\
	&=-m_b\pa{\epsilon^{bd}n_d}\nabla_a\pa{\epsilon^{ac}m_c}=-m_bm^b\nabla_a\pa{-n^a}\\
	&=\nabla_an^a
	=\kappa_n.
\end{align*}
\usetagform{default}\!\!
In conclusion,
\begin{align}
	K_{ab}=\kappa_nm_am_b.
\end{align}
We can now compute the commutator $\br{m,n}^a$:
\begin{align*}
	\br{m,n}^a&=m^b\nabla_bn^a-n^b\nabla_bm^a
	=\epsilon^{bc}n_c\nabla_bn^a-n^b\nabla_b\pa{\epsilon^{ac}n_c}\\
	&=\epsilon^{bc}n_c\pa{{K_b}^a+n_b\alpha^a}-\epsilon^{ac}\alpha_c
	=\epsilon_{bc}n^cK^{ba}+\epsilon^{bc}n_cn_b\alpha^a-\epsilon^{ac}\pa{\kappa_mm_c}\\
	&=m_bK^{ba}+0-\kappa_m\pa{\epsilon^{ac}m_c}
	=m_b\pa{\kappa_nm^bm^a}-\kappa_mn^a\\
	&=\kappa_nm^a-\kappa_mn^a.
\end{align*}
Knowing this, we can now show that
\usetagform{brackets}
\begin{align*}
	\L_m\L_n\log\ell&=\br{\L_m,\L_n}\log\ell+\L_n\L_m\log\ell
	=\L_{\br{m,n}}\log\ell-\L_n\kappa_m
	\tag{by Eq.~(\ref{eq:CurvatureLength})}\\
	&=\pa{\kappa_n\L_m\log\ell-\kappa_m\L_n\log\ell}-\L_n\kappa_m
	=\kappa_n\pa{-\kappa_m}-\kappa_m\L_n\log\ell-\L_n\kappa_m\\
	&=-\pa{\vec\L_n+\kappa_n+\L_n\log\ell}\kappa_m,
\end{align*}
\usetagform{default}\!\!
where the arrow on top of $\L_n$ indicates that it acts as a differential operator on any term outside the parentheses.  By acting with this operator again, we obtain the identity
\begin{align*}
	\L_m^2\L_n\log\ell&=\L_m\pa{\L_m\L_n\log\ell}=-\vec\L_m\pa{\vec\L_n+\kappa_n+\L_n\log\ell}\kappa_m\\
	&=-\vec\L_m\pa{\vec\L_n+\kappa_n}\kappa_m-\pa{\L_m\L_n\log\ell}\kappa_m-\pa{\L_m\kappa_m}\L_n\log\ell.
\end{align*}
In summary, we have obtained the following useful relations:
\begin{align}
	\L_T\phi&=\ell\sqrt{g^P}\L_m\phi,\\
	\kappa_n&=\nabla_an^a=g^{ab}\nabla_an_b=m^am^b\nabla_an_b,\\
	\kappa_m&=\nabla_am^a=\epsilon^{ab}\nabla_an_b=-n^am^b\nabla_an_b,\\
	\alpha^a&=n^b\nabla_bn^a=-\kappa_mm^a,\\
	K_{ab}&=\nabla_an_b-n_a\alpha_b=\kappa_nm_am_b,\\
	\br{m,n}^a&=\kappa_nm^a-\kappa_mn^a,\\
	\L_m\log\ell&=-\kappa_m,\\
\label{eq:LmLn}
	\L_m\L_n\log\ell&=-\pa{\vec\L_n+\kappa_n+\L_n\log\ell}\kappa_m,\\
	\L_m^2\L_n\log\ell&=-\vec\L_m\pa{\vec\L_n+\kappa_n}\kappa_m-\pa{\L_m\L_n\log\ell}\kappa_m-\pa{\L_m\kappa_m}\L_n\log\ell.
\end{align}

\subsection{Derivation of the geometric formulation}

We can now recast the terms entering the foliation condition \eqref{eq:FoliationCondition} in terms of the geometric invariants introduced in Eq.~\eqref{eq:GeometricInvariants}.  First, recall from Eqs.~\eqref{eq:GeneralA} and \eqref{eq:GeneralB} that
\begin{align}
	A&=-g^T\ab{\eta}^2\ell\pa{\kappa_n+\L_n\log\ab{\sqrt{-g^T}\ab{\eta}^2\ell}}+A_\Omega,\\
\label{eq:AOmega}
	A_\Omega&=-\frac{g^T\ell^2}{\alpha^2}\Omega'(u)\br{\Omega(u)-\Omega_Z},\\
	B&=-g^T\ab{\eta}^2\ell^2,
\end{align}
where $g^T=-\alpha^2\rho^2$ is the determinant of the toroidal metric.  We will assume for the moment that $\Omega(u)$ is a constant, in which case $A_\Omega=0$, and we can thus omit this term.  Then, we see that
\begin{align}
	\L_TB&=\ell\sqrt{g^P}\L_mB=\ell\sqrt{g^P}\L_m\pa{-g^T\ab{\eta}^2\ell^2}
	=\ell\sqrt{g^P}\pa{-g^T\ab{\eta}^2\ell^2}\L_m\log\ab{-g^T\ab{\eta}^2\ell^2}
	\nonumber\\
	&=\ell B\sqrt{g^P}\br{\L_m\log\ab{-g^T\ab{\eta}^2}+\L_m\log\ell^2}
	=\ell B\sqrt{g^P}\pa{\Delta_m+\kappa_m+2\L_m\log\ell}
	\nonumber\\
	&=\ell B\sqrt{g^P}\pa{\Delta_m-\kappa_m}.
	\nonumber
\end{align}
Proceeding in the same vein, we find that
\begin{align*}
	\L_T^2B&=\L_T\pa{\L_TB}
	=\ell\sqrt{g^P}\L_m\br{\ell B\sqrt{g^P}\pa{\Delta_m-\kappa_m}}\\
	&=\ell\sqrt{g^P}\Big[\L_m\pa{\ell}B\sqrt{g^P}\pa{\Delta_m-\kappa_m}+\ell\L_m\pa{B}\sqrt{g^P}\pa{\Delta_m-\kappa_m}\\
	&\qquad\qquad\qquad+\ell B\L_m\pa{\sqrt{g^P}}\pa{\Delta_m-\kappa_m}+\ell B\sqrt{g^P}\L_m\pa{\Delta_m-\kappa_m}\Big]\\
	&=\ell\sqrt{g^P}\Big[\pa{-\ell\kappa_m}B\sqrt{g^P}\pa{\Delta_m-\kappa_m}+\pa{\L_TB}\pa{\Delta_m-\kappa_m}\\
	&\qquad\qquad\qquad+\ell B\sqrt{g^P}\L_m\pa{\log\sqrt{g^P}}\pa{\Delta_m-\kappa_m}+\ell B\sqrt{g^P}\L_m\pa{\Delta_m-\kappa_m}\Big].
\end{align*}
After substituting the previous formula for $\L_TB$, this simplifies to
\begin{align*}
	\L_T^2B&=\ell^2Bg^P\br{-\kappa_m\pa{\Delta_m-\kappa_m}+\pa{\Delta_m-\kappa_m}^2+\L_m\log\sqrt{g^P}\pa{\Delta_m-\kappa_m}+\L_m\pa{\Delta_m-\kappa_m}}\\
	&=\ell^2Bg^P\pa{\vec\L_m+\Delta_m-2\kappa_m+\L_m\log\sqrt{g^P}}\pa{\Delta_m-\kappa_m}.
\end{align*}
Next, note that $A$ may be rewritten as
\begin{align}
	A=\frac{B}{\ell}\pa{\Delta_n+\L_n\log\ell}.
\end{align}
Hence,
\begin{align*}
	\L_TA&=\ell\sqrt{g^P}\L_mA=\ell\sqrt{g^P}\L_m\br{\frac{B}{\ell}\pa{\Delta_n+\L_n\log\ell}}\\
	&=B\sqrt{g^P}\L_m\pa{\Delta_n+\L_n\log\ell}+\pa{\Delta_n+\L_n\log\ell}\ell\sqrt{g^P}\L_m\pa{\frac{B}{\ell}}.
\end{align*}
Since
\begin{align*}
	\ell\sqrt{g^P}\L_m\pa{\frac{B}{\ell}}
	&=\sqrt{g^P}\L_mB+\ell\sqrt{g^P}B\L_m\pa{\frac{1}{\ell}}
	=\frac{1}{\ell}\L_TB-\frac{1}{\ell}\sqrt{g^P}B\L_m\ell\\
	&=B\sqrt{g^P}\pa{\Delta_m-\kappa_m}-\frac{1}{\ell}\sqrt{g^P}B\pa{-\ell\kappa_m}
	=B\sqrt{g^P}\Delta_m,
\end{align*}
it immediately follows that
\begin{align}
	\L_TA&=B\sqrt{g^P}\pa{\vec\L_m+\Delta_m}\pa{\Delta_n+\L_n\log\ell}.
\end{align}
Finally, we can compute
\begin{align*}
	\L_T^2A&=\L_T\pa{\L_TA}
	=\ell\sqrt{g^P}\L_m\br{B\sqrt{g^P}\frac{\L_TA}{B\sqrt{g^P}}}\\
	&=\ell\sqrt{g^P}\br{\L_m\pa{B}\sqrt{g^P}\frac{\L_TA}{B\sqrt{g^P}}+B\L_m\pa{\sqrt{g^P}}\frac{\L_TA}{B\sqrt{g^P}}+B\sqrt{g^P}\L_m\pa{\frac{\L_TA}{B\sqrt{g^P}}}} .
\end{align*}
Since
\begin{align*}
	\L_mB=\frac{1}{\ell\sqrt{g^P}}\L_TB
	=\frac{1}{\ell\sqrt{g^P}}\br{\ell B\sqrt{g^P}\pa{\Delta_m-\kappa_m}}
	=B\pa{\Delta_m-\kappa_m},
\end{align*}
we can factorize the previous expression as
\begin{align*}
	\L_T^2A&=\ell Bg^P\br{\pa{\Delta_m-\kappa_m}\frac{\L_TA}{B\sqrt{g^P}}+\L_m\pa{\log\sqrt{g^P}}\frac{\L_TA}{B\sqrt{g^P}}+\L_m\pa{\frac{\L_TA}{B\sqrt{g^P}}}}\\
	&=\ell Bg^P\pa{\vec\L_m+\Delta_m-\kappa_m+\L_m\log\sqrt{g^P}}\pa{\frac{\L_TA}{B\sqrt{g^P}}}.
\end{align*}
In summary, we have shown that
\begin{align}
	\L_TA&=B\sqrt{g^P}\pa{\vec\L_m+\Delta_m}\pa{\Delta_n+\L_n\log\ell},\\
	\L_TB&=\ell B\sqrt{g^P}\pa{\Delta_m-\kappa_m},\\
	\L_T^2A&=\ell Bg^P\pa{\vec\L_m+\Delta_m-\kappa_m+\L_m\log\sqrt{g^P}}\pa{\frac{\L_TA}{B\sqrt{g^P}}},\\
	\L_T^2B&=\ell^2Bg^P\pa{\vec\L_m+\Delta_m-2\kappa_m+\L_m\log\sqrt{g^P}}\pa{\Delta_m-\kappa_m}.
\end{align}
Next, recall that the foliation condition \eqref{eq:FoliationCondition} can be written as $\det M=0$, where the matrix $M$ is 
\begin{align}
	M=\begin{bmatrix}
		\L_TA & \L_TB \\
		\L_T^2A & \L_T^2B
	\end{bmatrix}.
\end{align}
We are thus free to replace the foliation condition by a new equation
\begin{align}
	\det\tilde{M}=0,
\end{align}
where $\tilde{M}$ can be taken to be any matrix whose determinant is proportional to that of $M$,
\begin{align}
\label{eq:MatrixDeterminant}
	\det M=\sigma\det\tilde{M},\qquad\sigma\neq0.
\end{align}
The simplest choice we could find is
\begin{align}
\label{eq:tildeM}
	\tilde{M}=
	\begin{bmatrix}
		X && Y \\
		\pa{\vec\L_m+\Delta_m}X && \L_mY
	\end{bmatrix},
\end{align}
where
\begin{align}
\label{eq:X}
	X&=\pa{\vec\L_m+\kappa_m}\Delta_n-\pa{\vec\L_n+\kappa_n}\kappa_m=\nabla_a\br{\Delta_nm^a-\kappa_mn^a},\\
\label{eq:Y}
	Y&=\Delta_m-\kappa_m,\\
\label{eq:sigma}
	\sigma&=\ell^2B^2\pa{g^P}^{3/2}.
\end{align}

\subsection{Derivation of the determinant form}

To obtain $\tilde{M}$ starting from $M$, we apply a sequence of transformations that leave the determinant unchanged.  First, following Eq.~\eqref{eq:MatrixDeterminant}, we note that we can strip from $\det M$ an overall factor of $\sigma=\ell^2B^2\pa{g^P}^{3/2}$ [hence, Eq.~\eqref{eq:sigma}], leaving the nontrivial part
\begin{align*}
	\bar{M}=
	\begin{bmatrix}
		\bar{M}_{11} && \Delta_m-\kappa_m \\
		\pa{\vec\L_m+\Delta_m-\kappa_m+\L_m\log\sqrt{g^P}}\bar{M}_{11} && \pa{\vec\L_m+\Delta_m-2\kappa_m+\L_m\log\sqrt{g^P}}\pa{\Delta_m-\kappa_m}
	\end{bmatrix},
\end{align*}
where the matrix entry $\bar{M}_{11}=\pa{\vec\L_m+\Delta_m}\pa{\Delta_n+\L_n\log\ell}$.  In terms of $Y=\Delta_m-\kappa_m$, this is just
\begin{align*}
	\bar{M}=\begin{bmatrix}
		\bar{M}_{11} && Y \\
		\pa{\vec\L_m+\Delta_m-\kappa_m+\L_m\log\sqrt{g^P}}\bar{M}_{11} && \pa{\vec\L_m+\Delta_m-2\kappa_m+\L_m\log\sqrt{g^P}}Y
	\end{bmatrix}.
\end{align*}
Next, we define a new matrix $\bar{\bar{M}}$ by multiplying $\bar{M}$ with a matrix $O$ of unit determinant,
\begin{align*}
	\bar{\bar{M}}=O\bar{M},\qquad O=
	\begin{bmatrix}
		1 && 0 \\
		\kappa_m-\Delta_m-\L_m\log\sqrt{g^P} && 1
	\end{bmatrix},
\end{align*}
so that $\det\bar{\bar{M}}=\det\bar{M}=\sigma^{-1}\det M$ still encodes the foliation condition.  The result is
\begin{align*}
	\bar{\bar{M}}=\begin{bmatrix}
		\bar{M}_{11} && Y \\
		\bar{\bar{M}}_{21} && \bar{\bar{M}}_{22}
	\end{bmatrix},
\end{align*}
where the new matrix entries are
\begin{align*}
	\bar{\bar{M}}_{21}&=\pa{\kappa_m-\Delta_m-\L_m\log\sqrt{g^P}}\bar{M}_{11}+\pa{\vec\L_m+\Delta_m-\kappa_m+\L_m\log\sqrt{g^P}}\bar{M}_{11}\\
	&=\L_m\bar{M}_{11}
	=\vec\L_m\pa{\vec\L_m+\Delta_m}\pa{\Delta_n+\L_n\log\ell},\\
	\bar{\bar{M}}_{22}&=\pa{\kappa_m-\Delta_m-\L_m\log\sqrt{g^P}}Y+\pa{+\vec\L_m\Delta_m-2\kappa_m+\L_m\log\sqrt{g^P}}Y\\
	&=\pa{\vec\L_m-\kappa_m}Y
	=\pa{\vec\L_m-\kappa_m}\pa{\Delta_m-\kappa_m}.
\end{align*}
Thus, we can simplify $\bar{\bar{M}}$ to
\begin{align*}
	\bar{\bar{M}}=\begin{bmatrix}
		\bar{M}_{11} && Y \\
		\L_m\bar{M}_{11} && \pa{\vec\L_m-\kappa_m}Y
	\end{bmatrix}.
\end{align*}
Note that at this point, the foliation condition in the form of the determinant of $\bar{\bar{M}}$ is manifestly independent of the poloidal metric $g^P$, as it should be (because only the foliation should matter).  In order to proceed, we must now expand
\begin{align*}
	\bar{M}_{11}=\pa{\vec\L_m+\Delta_m}\pa{\Delta_n+\L_n\log\ell}
	=\pa{\vec\L_m+\Delta_m}\Delta_n+\Delta_m\L_n\log\ell+\L_m\L_n\log\ell.
\end{align*}
The last term, $\L_m\L_n\log\ell$, can be simplified using Eq.~\eqref{eq:LmLn}, leading to
\begin{align*}
	\bar{M}_{11}&=\pa{\vec\L_m+\Delta_m}\Delta_n-\pa{\vec\L_n+\kappa_n}\kappa_m+\pa{\Delta_m-\kappa_m}\L_n\log\ell\\
	&=X+Y\Delta_n+Y\L_n\log\ell,
\end{align*}
where in the last step we used the definitions \eqref{eq:X} and \eqref{eq:Y} of $X$ and $Y$.  It now results that
\begin{align*}
	\bar{\bar{M}}=\begin{bmatrix}
		X+Y\pa{\Delta_n+\L_n\log\ell} && Y \\
		\L_m\br{X+Y\pa{\Delta_n+\L_n\log\ell}} && \pa{\vec\L_m-\kappa_m}Y
	\end{bmatrix},
\end{align*}
which has determinant
\begin{align*}
	\det\bar{\bar{M}}&=\br{X+Y\pa{\Delta_n+\L_n\log\ell}}\pa{\vec\L_m-\kappa_m}Y-Y\L_m\br{X+Y\pa{\Delta_n+\L_n\log\ell}}\\
	&=X\L_mY-Y\L_mX-XY\kappa_m+Y\pa{\Delta_n+\L_n\log\ell}\pa{\vec\L_m-\kappa_m}Y-Y\L_m\br{Y\pa{\Delta_n+\L_n\log\ell}}\\
	&=X\L_mY-Y\L_mX-XY\kappa_m+Y\Delta_n\L_mY-Y\Delta_n\kappa_mY+Y\L_n\log\ell\pa{\vec\L_m-\kappa_m}Y\\
	&\qquad-Y\pa{\Delta_n+\L_n\log\ell}\L_mY-Y^2\L_m\pa{\Delta_n+\L_n\log\ell}\\
	&=X\L_mY-Y\L_mX-XY\kappa_m-Y\Delta_n\kappa_mY+Y\L_n\log\ell\pa{-\kappa_m}Y-Y^2\L_m\pa{\Delta_n+\L_n\log\ell}\\
	&=X\L_mY-Y\L_mX-XY\kappa_m+Y^2\br{-\kappa_m\pa{\Delta_n+\L_n\log\ell}-\L_m\pa{\Delta_n+\L_n\log\ell}}.
\end{align*}
Now observe that
\begin{align}
	XY\kappa_m=XY\pa{\Delta_m+\kappa_m-\Delta_m}=XY\Delta_m-XY^2,
\end{align}
and hence, that
\begin{align*}
	\det\bar{\bar{M}}=X\L_mY-Y\L_mX-XY\Delta_m+Y^2\br{X-\kappa_m\pa{\Delta_n+\L_n\log\ell}-\L_m\pa{\Delta_n+\L_n\log\ell}}.
\end{align*}
The term in brackets vanishes:
\begin{align*}
	X&-\kappa_m\pa{\Delta_n+\L_n\log\ell}-\L_m\pa{\Delta_n+\L_n\log\ell}\\
	=X&-\kappa_m\pa{\Delta_n+\L_n\log\ell}-\L_m\Delta_n-\L_m\L_n\log\ell\\
	=X&-\kappa_m\pa{\Delta_n+\L_n\log\ell}-\L_m\Delta_n+\pa{\vec\L_n+\kappa_n+\L_n\log\ell}\kappa_m\\
	=X&-\br{\pa{\vec\L_m+\kappa_m}\Delta_n-\pa{\vec\L_n+\kappa_n}\kappa_m}
	=0.
\end{align*}
In conclusion, we have found that
\begin{align*}
	\det\bar{\bar{M}}=X\L_mY-Y\L_mX-XY\Delta_m.
\end{align*}
By Eq.~\eqref{eq:MatrixDeterminant}, this proves the claim \eqref{eq:tildeM} that
\begin{align*}
	\det\tilde M=X\L_mY-Y\L_mX-XY\Delta_m
	=\det\bar{\bar{M}}
	=\det\bar{M}
	=\sigma^{-1}\det M
\end{align*}
still encodes the foliation condition.

\subsection{The case of nonconstant field line angular velocity}

So far, we have assumed that $\Omega(u)$ is a constant, which made $A_\Omega$ vanish.  We now generalize to the case of $\Omega(u)$ nonconstant, and consequently reintroduce $A_\Omega$.  The foliation condition
$\det M=0$ is then modified to
\begin{align}
	\det\pa{M+M_\Omega}=\det M+\det M_\Omega=0,
\end{align}
where
\begin{align}
	M_\Omega=
	\begin{bmatrix}
		\L_TA_\Omega & \L_TB \\
		\L_T^2A_\Omega & \L_T^2B
	\end{bmatrix}.
\end{align}
Recalling the definition \eqref{eq:AOmega} of $A_\Omega$ and the fact that $\L_mf(u)=0$ for any function $f$, we see that
\begin{align*}
	\L_TA_\Omega&=\ell\sqrt{g^P}\L_mA_\Omega
	=\ell\sqrt{g^P}\L_m\pa{-\frac{g^T\ell^2}{\alpha^2}\Omega'(u)\br{\Omega(u)-\Omega_Z}}\\
	&=\ell\sqrt{g^P}\L_m\pa{-\frac{g^T\ell^2}{\alpha^2}}\Omega'(u)\br{\Omega(u)-\Omega_Z}-\ell\sqrt{g^P}\pa{-\frac{g^T\ell^2}{\alpha^2}}\Omega'(u)\L_m\Omega_Z\\
	&=\frac{\ell^3g^T\sqrt{g^P}}{\alpha^2}\Omega'(u)\pa{-\br{\Omega(u)-\Omega_Z}\L_m\log\ab{-\frac{g^T\ell^2}{\alpha^2}}+\L_m\Omega_Z}.
\end{align*}
Using the invariants the defined in Eq.~\eqref{eq:GeometricInvariants}, note that
\begin{align*}
	\L_m\log\ab{\frac{-g^T\ell^2}{\alpha^2}}
	&=\L_m\log\ab{\frac{-g^T\ab{\eta}^2\ell^2}{\alpha^2\ab{\eta}^2}}
	=\L_m\log\ab{-g^T\ab{\eta}^2}-\L_m\log\ab{\alpha^2\ab{\eta}^2}+2\L_m\log\ell\\
	&=\Delta_m+\kappa_m-\L_m\log\ab{\alpha^2\ab{\eta}^2}-2\kappa_m
	=\Delta_m-\kappa_m-\alpha_m.
\end{align*}
Hence,
\begin{align}
	\L_TA_\Omega=\frac{\ell^3g^T\sqrt{g^P}}{\alpha^2}\Omega'(u)\cu{-\br{\Omega(u)-\Omega_Z}\pa{\Delta_m-\kappa_m-\alpha_m}+\L_m\Omega_Z}.
\end{align}
We now define
\begin{align}
	X_\Omega=\frac{\ell}{\alpha^2\ab{\eta}^2}\Omega'(u)\cu{\br{\Omega(u)-\Omega_Z}\pa{\Delta_m-\kappa_m-\alpha_m}-\L_m\Omega_Z}.
\end{align}
This quantity is invariant under changes of the foliation.  The overall prefactor in $X_\Omega$ is chosen for later convenience.  In terms of this new quantity, we have
\begin{align}
	\L_TA_\Omega=-\ell^2g^T\sqrt{g^P}\ab{\eta}^2X_\Omega,
\end{align}
and thus,
\begin{align*}
	\L_T^2A_\Omega&=\L_T\pa{\L_TA_\Omega}
	=\ell\sqrt{g^P}\L_m\pa{\L_TA_\Omega}
	=-\ell\sqrt{g^P}\L_m\pa{\ell^2g^T\sqrt{g^P}\ab{\eta}^2X_\Omega}\\
	&=-\ell^3g^Tg^P\ab{\eta}^2\br{\L_m\log\ab{\ell^2g^T\sqrt{g^P}\ab{\eta}^2}X_\Omega+\L_mX_\Omega}.
\end{align*}
As before, we can expand
\begin{align*}
	\L_m\log\ab{\ell^2g^T\sqrt{g^P}\ab{\eta}^2}
	&=2\L_m\log\ell+\L_m\log\ab{g^T\ab{\eta}^2}+\L_m\log\sqrt{g^P}\\
	&=-2\kappa_m+\Delta_m+\kappa_m+\L_m\log\sqrt{g^P}
	=\Delta_m-\kappa_m +\L_m\log\sqrt{g^P},
\end{align*}
from which it follows that
\begin{align}
	\L_T^2A_\Omega=-\ell^3g^Tg^P\ab{\eta}^2\br{\pa{\Delta_m-\kappa_m+\L_m\log\sqrt{g^P}}X_\Omega+\L_mX_\Omega}.
\end{align}
In summary, we have established that
\begin{align}
	\L_TA_\Omega&=-\ell^2g^T\sqrt{g^P}\ab{\eta}^2X_\Omega,\\
	\L_TB&=\ell B\sqrt{g^P}\pa{\Delta_m-\kappa_m},\\
	\L_T^2A_\Omega&=-\ell^3g^Tg^P\ab{\eta}^2\br{\pa{\Sigma_m+\kappa_m}X_\Omega+\L_mX_\Omega},\\
	\L_T^2B&=\ell^2Bg^P\pa{\Sigma_m+\vec\L_m}\pa{\Delta_m-\kappa_m},\\
	\Sigma_m&=\Delta_m-2\kappa_m+\L_m\log\sqrt{g^P}.
\end{align}
Once again following Eq.~\eqref{eq:MatrixDeterminant}, we note that we can strip from $\det M_\Omega$ an overall factor of $\sigma=-\ell^4Bg^T\pa{g^P}^{3/2}\ab{\eta}^2=\ell^2B^2\pa{g^P}^{3/2}$ [in agreement with the $\sigma$ defined in Eq.~\eqref{eq:sigma}], leaving the nontrivial part
\begin{align*}
	\bar{M}_\Omega&=
	\begin{bmatrix}
		X_\Omega && Y \\
		\pa{\Sigma_m+\kappa_m}X_\Omega+\L_mX_\Omega && \pa{\Sigma_m+\vec\L_m}Y
	\end{bmatrix},
\end{align*}
where $Y=\Delta_m-\kappa_m$ as usual.  The matrix $\bar{M}_\Omega$ has determinant
\begin{align*}
	\det\bar{M}_\Omega
	=\sigma^{-1}\det M_\Omega
	&=X_\Omega\pa{\Sigma_m+\vec\L_m}Y-Y\pa{\Sigma_m+\kappa_m}X_\Omega-Y\L_mX_\Omega\\
	&=X_\Omega \L_m Y-Y(\vec{\L}_m+ \kappa_m)X_\Omega
	=\det\tilde{M}_\Omega,
\end{align*}
where in the last step, we defined the final matrix
\begin{align}
\label{eq:tildeMOmega}
	\tilde{M}_\Omega=
	\begin{bmatrix}
		X_\Omega && Y \\
		\pa{\vec\L_m+\kappa_m}X_\Omega && \L_mY
	\end{bmatrix}.
\end{align}
Finally, since the sum of the determinants of both matrices \eqref{eq:tildeM} and \eqref{eq:tildeMOmega} equals the original determinant $\det\pa{M+M_\Omega}$ up to the overall common scaling factor $\sigma$, one can rewrite the complete geometric foliation condition as Eq.~\eqref{eq:GeometricCondition}.  This completes our derivation.\hfill\includegraphics[scale=.008]{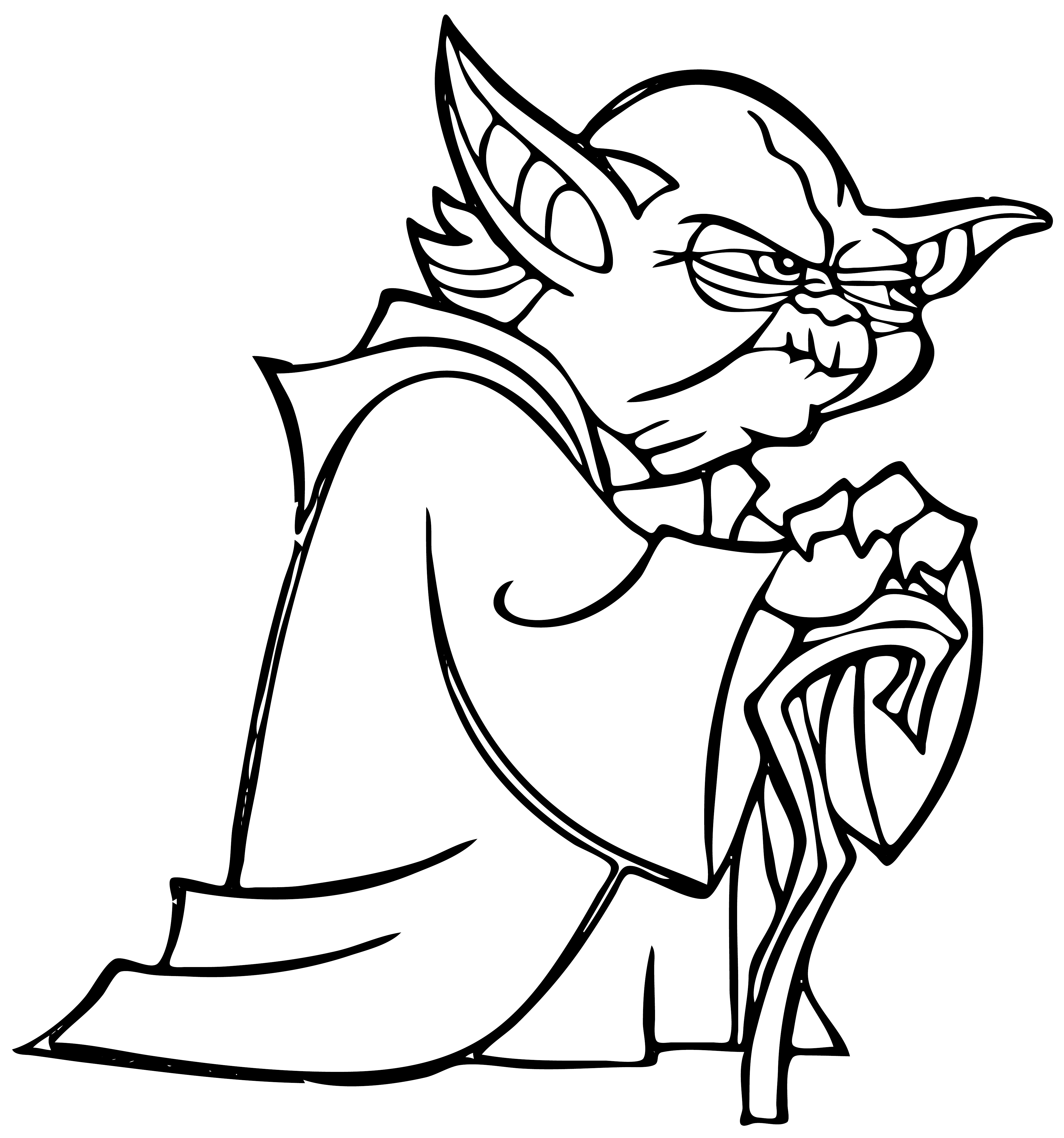}

\bibliographystyle{utphys}
\bibliography{FFF}

\end{document}